\documentclass[final,1p,times,twocolumn,authoryear]{elsarticle}
\usepackage{amssymb}
\journal{}
\usepackage{amsmath}
\usepackage{graphicx}

\begin{document}

\begin{frontmatter}



\title{Formation of working memory in a spiking neuron network accompanied by astrocytes}

\author[mymainaddress,mysecondaryaddress]{Susanna Yu. Gordleeva\corref{mycorrespondingauthor}}

\author[mymainaddress]{Yulia A. Tsybina}
\author[mymainaddress]{Mikhail I. Krivonosov}
\author[mymainaddress]{Mikhail V. Ivanchenko}
\author[mymainaddress,mythirdaddress,myfourthaddress]{Alexey A. Zaikin}
\author[mymainaddress,mysecondaryaddress]{Victor B. Kazantsev}
\author[mymainaddress,myfifthaddress]{Alexander N. Gorban}
\cortext[mycorrespondingauthor]{Corresponding author}

\address[mymainaddress]{Lobachevsky State University of Nizhny Novgorod, Nizhny Novgorod, Russia}
\address[mysecondaryaddress]{Neuroscience and Cognitive Technology Laboratory, Center for Technologies in Robotics and Mechatronics Components, Innopolis University, Innopolis, Russia}
\address[mythirdaddress]{Centre for Analysis of Complex Systems, Sechenov First Moscow State Medical University (Sechenov University), Moscow, Russia}
\address[myfourthaddress]{Institute for Women’s Health and Department of Mathematics, University College London, London, UK}
\address[myfifthaddress]{University of Leicester, UK}

\begin{abstract}
We propose a biologically plausible computational model of working memory (WM) implemented by the spiking neuron network (SNN) interacting with a network of astrocytes. SNN is modelled by the synaptically coupled Izhikevich neurons with a non-specific architecture connection topology. Astrocytes generating calcium signals are connected by local gap junction diffusive couplings and interact with neurons by chemicals diffused in the extracellular space. Calcium elevations occur in response to the increase of concentration of a neurotransmitter released by spiking neurons when a group of them fire coherently. In turn, gliotransmitters are released by activated astrocytes modulating the strengths of synaptic connections in the corresponding neuronal group. Input information is encoded as two-dimensional patterns of short applied current pulses stimulating neurons. The output is taken from frequencies of transient discharges of corresponding neurons. We show how a set of information patterns with quite significant overlapping areas can be uploaded into the neuron-astrocyte network and stored for several seconds. Information retrieval is organised by the application of a cue pattern representing the one from the memory set distorted by noise. We found that successful retrieval with level of the correlation between recalled pattern and ideal pattern more than 90\% is possible for multi-item WM task. Having analysed the dynamical mechanism of WM formation, we discovered that astrocytes operating at a time scale of a dozen of seconds can successfully store traces of neuronal activations corresponding to information patterns. In the retrieval stage, the astrocytic network selectively modulates synaptic connections in SNN leading to the successful recall. Information and dynamical characteristics of the proposed WM model agrees with classical concepts and other WM models.
\end{abstract}

\begin{keyword}
spiking neural network \sep astrocyte \sep neuron-astrocyte interaction \sep working memory 
\end{keyword}
\end{frontmatter}

\section{Introduction}

In neuroscience, the understanding of the functional role of astrocytes in CNS is still open to debate \citep{Savtchouk2018}, but now there is much evidence demonstrating the involvement of astrocytes in local synaptic plasticity and coordination of network activity \citep{Durkee2019}, and as a result in information processing and memory encoding \citep{Santello2019}. Astrocytes sense synaptic activity and respond to it with the transient elevation of the intracellular Ca$^{2+}$ concentration (lasting from hundreds of a millisecond to a dozen of seconds). Such Ca$^{2+}$ signals in astrocytes have been observed in different brain regions and also in the cortex, appearing there in response to the mechanic sensory stimulation \citep{Stobart2018,Wang2006_2,Takata2011} and visual sensory stimulation \citep{Chen2012,Schummers2008,Perea2014}. Ca$^{2+}$ activation can trigger the release of gliotransmitters from astrocyte, which in turn affect the dynamics of presynaptic and postsynaptic terminals resulting in modulations of synaptic transmission \citep{Araque2014}. The gliotransmitter-mediated synaptic modulation lasts from a dozen of seconds \citep{Jourdain2007,Perea2014} to a dozen of minutes \citep{Perea2007,Navarrete2012,Stellwagen2006} contributing to both short- and long-term synaptic plasticity. 
Obviously, there is a qualitative coincidence of time scales of astrocyte-mediated synaptic modulation with the working memory (WM) timings during decision making. Based on this and the other following facts of astrocyte participation in neuronal signalling, we hypothesized that the astrocytes may be involved in the WM formation. In particular, recent in vivo studies have shown the participation of astrocytes in synchronization of certain cortical network activity \citep{Takata2011,Chen2012,Perea2014,Paukert2014}, cognitive functions and behaviour \citep{Sardinha2017,Poskanzer2016}. Experimental evidence shows that astrocyte pathology in medium PFC impairs the WM and learning functions \citep{Lima2014}, increasing of astrocyte density  enhances short-term memory performance \citep{DeLuca2020}, recognition memory performance and disruption of WM depend on the gliotransmitter release from astrocyte in the hippocampus \citep{Robin2018,Han2012}. Despite these numerous experimental insights of the contribution of astrocytes to synaptic modulations in neuronal signalling, the possible role of astrocyte in information processing and learning is still a subject of discussion  \citep{Kastanenka2019,Kanakov2019}.

Considering the significance of WM processes, the challenge of finding alternative mechanisms and the experimental evidence of the astrocytic role in information processing in CNS, it is interesting to study astrocyte-induced modulation of synaptic transmission in the WM organisation. Specifically, we assume that the NMDAR-mediated potentiation of excitatory synapses induced by the D-serine released from astrocyte in PFC \citep{Fossat2011,Takata2011,Chen2012} plays an essential role in WM. To test this hypothesis, we developed a novel neuron-astrocyte network model for visual WM to reflect experimental data on the structure, connectivity, and neurophysiology of the neuron-astrocytic interaction in underlying cortical tissue. We focused on the implementation of a multi-item WM task in DMS framework representing classical neuropsychological paradigm \citep{Miller1996}, which was previously studied using a recurrent neural network \citep{Brunel2001,Amit2003,Amit2013,Fiebig2016}. In our model, memory is associated with item-specific patterns of astrocyte-induced enhancement of excitatory synaptic transmission. We show how the biologically relevant neuron-astrocyte network model implements loading, storage and cued retrieval of multiple items with significant overlapping. The memory items are encoded in neuronal populations in the form of discrete high-frequency bursts rather than persistent spiking.

In the following, we review some related works (Section 2), describe the proposed model and methods in detail (Section 3), present the results (Section 4), and finally, we conclude this work in Section 5.

\section{Related work}\label{sec_review}
 The concept of WM proposes the ability to temporarily store and process information in goal-directed behaviour. WM is crucial in the generation of higher cognitive functions for
 both humans and other animals  \citep{Baddeley1986,Conway2003,Baddeley2012}. In primates, visual WM has been studied in delay tasks, such as 
 delayed matching to sample (DMS), which 
 require memory to be held during a brief delay period lasting for several seconds \citep{Miller1996}. Recordings in the monkeys' prefrontal cortices (PFCs) during the delay task showed that some neurons displayed persistent and stimulus-specific delay-period activity \citep{Fuster1971,Funahashi1989,Barak2010,Shafi2007,Funahashi2017}. Delay persistent activity is considered the neural correlate of WM \citep{GoldmanRakic1995,Constantinidis2018}. 

The classical theoretical memory models suggest that an information item can be stored with sustained neural activity which emerges via activation of stable activity patterns in the network (e.g. attractors) \citep{Hopfield1982,Amit1995,Wang2001,Wimmer2014} recently reviewed by \citet{Zylberberg2017,Chaudhuri2016}. These WM models propose that the generation of the persistent activity can be the result of an intrinsic property of the neurons (including the generation of the bistability mediated by the voltage-gated inward currents \citep{Kass2005} and Ca$^{2+}$-triggered long-term changes in neuronal excitability \citep{Fransn2006}) and can be induced by the connectivity within the neural circuit with feed-forward \citep{Ganguli2009,Goldman2009} or recurrent \citep{Kilpatrick2013,Koulakov2002} architecture. In such models, memory recall is impossible from a silent inactive state. For many WM models of persistent activity based on recurrent connectivity, small deviations in the network structure destroy the persistence. Moreover, a spiking form of storage information is energetically unfavourable because of the high metabolic value of action potentials \citep{Attwell2001}.

In theoretical studies a concept of oscillatory sub-cycles storing $7 \pm 2$ in oscillatory neuronal networks were proposed by \citet{Lisman1995}. Other models employ oscillatory activity of spiking neuron networks with afterdepolarization to memorize set of information patterns at different phases of rhythmic oscillations \citep{KLINSHOV2008,Borisyuk2013}.

Recently, the persistent activity hypothesis has been undergoing critical reviews  \citep{Lundqvist2018} based on the experimental findings in the rodents, rats, and primates showing that the robust persistent activity does not last for the entire delay period, but rather sequential neuronal firing is observed during the delay period suggesting that the PFC neural network may support WM based on dynamically-changing neuronal activity \citep{Fujisawa2008,Runyan2017,Park2019,Ozdemir2020,Lundqvist2016}. Despite the considerable progress which has been made in identifying the neurophysiological mechanisms contributing to WM in mammals \citep{DEsposito2015,Zylberberg2017}, the ongoing debate focuses on the generation mechanisms of the delay period activity that appears to underlie the WM  \citep{Constantinidis2018,Sreenivasan2019}. 
Currently, one of the recognised experimentally based hypotheses of the WM mechanism underlining the delay activity (not necessarily persistent) is the synaptic plasticity in the PFC \citep{Tsodyks1997,Wang2006,Hempel2000,Erickson2010}. Synaptic plasticity implies a rapid regulation of the strengths of individual synapses in response to specific patterns of correlated synaptic activity and contributes to the activity-dependent refinement of neural circuitry. Following these findings, alternative synaptic-based WM models have been proposed \citep{Mongillo2008,Manohar2019,Barak2014,Koutsikou2018}. In these models, memory items are stored by stimulus-specific patterns of synaptic facilitation in neuronal circuit. Synaptic plasticity does not require neurons to show a persistent activity for the entire period of the memory task, which results in a robust and more metabolically efficient mechanism. Some synaptic WM models based on short-term non-associative synaptic facilitation  \citep{Mongillo2008,Lundqvist2011,Mi2017} allow reading out and refreshing existing representations maintained in the synaptic structure. Others have proposed fast Hebbian activity-dependent synaptic plasticity \citep{Fiebig2016,Sandberg2003} for encoding and maintenance of novel associations. 

There are a few attempts to investigate theoretically the role of astrocyte-induced modulation of synaptic transmission in memory formation. \citet{Shen2007} demonstrate one of  the first results of simulating the coupling of Hopfield neural network, astrocyte, and cerebrovascular activity. Although there remained a far stretch to a biophysical model, the resulted suggested that a modification of the synapse strengths allows the neuronal firing and the cerebrovascular flow to be compatible on a meso-scale; with astrocyte signalling added, limit cycles exist in the coupled networks. \citet{Tewari2013} and \citet{Wade2011} study how bidirectional coupling between astrocytes and
neurons in small neuron-astrocyte ensembles mediates learning and dynamic coordination in the brain. Recent interesting theoretical study proposes a self-repairing spiking astrocyte-neural network combined with novel learning rule based on the spike-timing-dependent plasticity and Bienenstock, Cooper, and Munro learning rule \citep{Liu2019}.

\section{Materials and methods}\label{sec_model}
\subsection{Neuron-astrocyte network model}
\begin{figure}
\centering
\includegraphics[width=1\columnwidth]{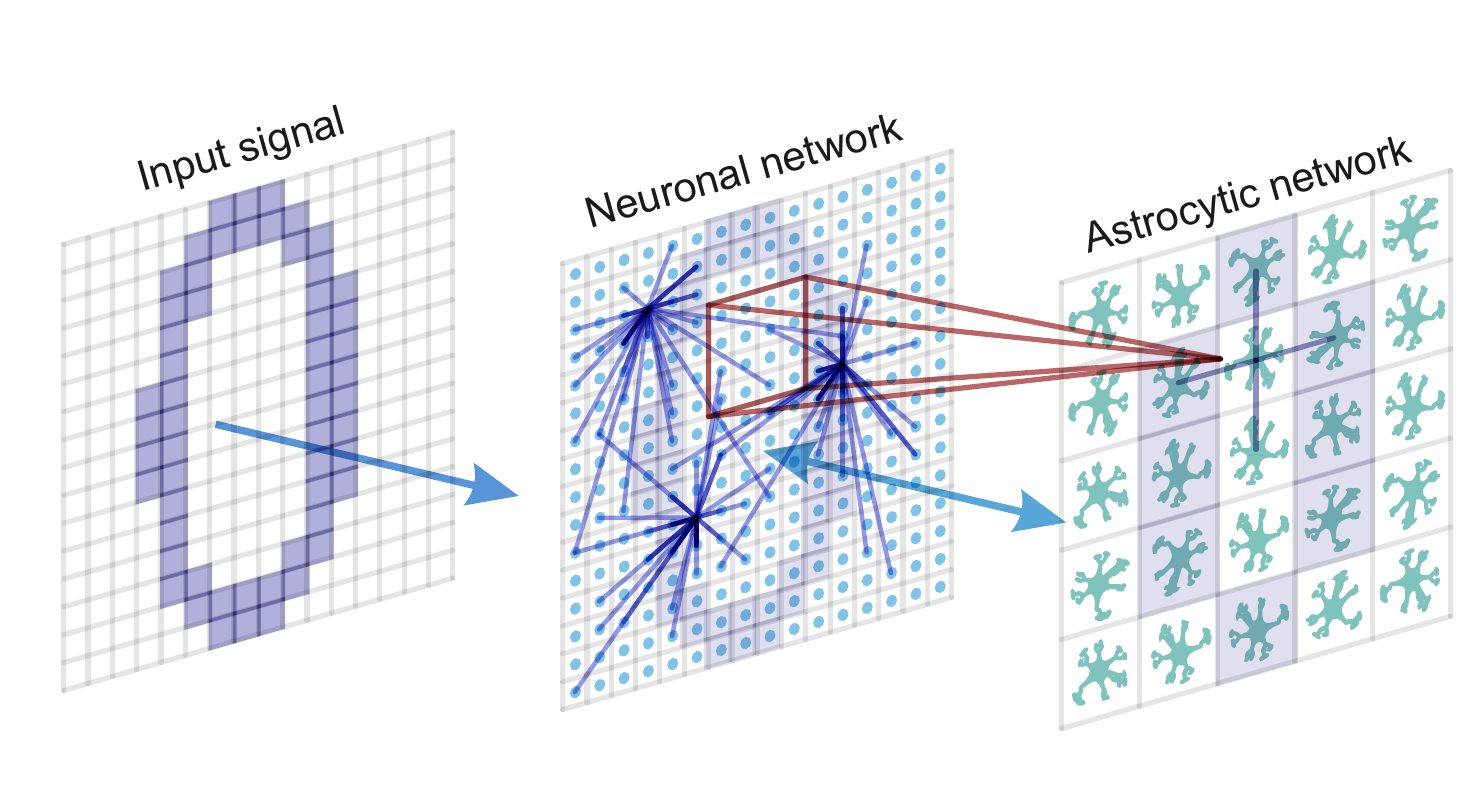}
\caption{Neuron-astrocyte network topology. Neuron-astrocyte network consists of two interacting layers: neural network layer and astrocytic layer. Neuronal network with dimension $W\times H$ ($79\times 79$) consists of synaptically coupled excitatory neurons modelled by the Izhikevich neuron. Neurons in network are connected randomly. The astrocytic network consists of diffusely connected astrocytes with dimension  $M\times N$ ($26\times 26$). Blue lines show connections between elements in each layer. We consider the bidirectional interaction between the neuronal and astrocytic layers. Each astrocyte is interconnected with a neuronal ensemble of $N_a=16$ neurons with dimensions $4\times 4$ with overlapping in one row (red lines). The input signal is fed to the neural network.}\label{fig_topo}
\end{figure}

Even though the balance of inhibition and excitation play was shown to play an important role in WM stabilization and to influence the WM capacity \citep{Barak2014}, we focus on the properties of astrocyte-induced modulation of excitatory synaptic transmission in the PFC. We take spiking neuronal network with dimension $W\times H$ consisting of synaptically coupled excitatory neurons formed by the Izhikevich model \citep{Izhikevich2003}. Neurons in the network are connected randomly with the connection length determined by the exponential distribution. 
It has been experimentally estimated that there is a little overlap in the spatial territories occupied by individual astrocytes in the cortex  \citep{Halassa2007}. An individual cortical astrocyte contacts on average 4-8 neuronal somata and 300–600 neuronal dendrites \citep{Halassa2007}. A cortical astrocyte has a “bushy”
ramified structure of the fine perisynaptic processes, which cover most of neuronal membranes within their reach \citep{Allen2017}. This allows the astrocyte to integrate and coordinate a unique volume of synaptic activity. Following the experimental data, astrocytic network compartment of our model is organised as a two-dimensional square lattice with only nearest-neighbour connectivity. Each astrocyte interacts with the neuronal ensemble of $N_a$ neurons with small overlapping. We consider the bidirectional communication between neuronal and astrocytic networks. The scheme of the network topology is shown in Fig.~\ref{fig_topo}.

Model equations are integrated using the Runge-Kutta fourth-order method with a fixed time step, $\Delta t = 0.1$ ms. A detailed listing of model parameters and values can be found in Tables 1 (neural network model), 2 (astrocytic network parameters), 3 (neuron-astrocytic interaction parameters) and 4 (stimulation and recall testing). The code is available at the link https://github.com/altergot/neuro-astro-network.

\subsection{Neuronal network}\label{sec_neur_model}
 Dynamics of the membrane potential of a single neuron is described by the Izhikevich model \citep{Izhikevich2003}:
\begin{equation}
\label{eq:Izh_main}
\begin{aligned}
&\frac{dV^{(i,j)}}{dt} = 0.04V^{(i,j)^{(2)}} + 5V^{(i,j)} - U^{(i,j)} + 140 +I^{(i,j)}_{\text{app}} + I^{(i,j)}_\text{syn}; \\
&\frac{dU^{(i,j)}}{dt} = a ( b V^{(i,j)} - U^{(i,j)});\\
\end{aligned}
\end{equation}
with the auxiliary after-spike resetting
\begin{equation}
    \label{eq:Izh_cond}
    \text{if } V^{(i,j)} \ge \text{30 mV, then}
        \begin{cases}
            V^{(i,j)}\gets c \\U^{(i,j)} \gets U^{(i,j)} + d,\\
        \end{cases}
\end{equation}
where the superscripts ($i = 1,\ldots, 79$, $j = 1,\ldots, 79$) correspond to a neuronal index, the transmembrane potential $V$ is given in
mV and time $t$ in ms. The applied currents $I^{(i,j)}_{\text{app}}$ simulate the input signal. Neurons receive a number of synaptic currents from other presynaptic neurons in the network, $N^{(i,j)}_{in}$, which are summed at the membrane according to the following equation \citep{Kazantsev2011,Esir2018}:
\begin{equation}
    \label{eq:I_syn}
    \begin{aligned}
    &I^{(i,j)}_{\text{syn}} =\sum_{k=1}^{N^{(i,j)}_{in}} \frac{g^{(i,j)}_{\text{syn}}(E_{\text{syn}}-V^{(i,j)})}{1+\exp(\frac{-V^{k}_{\text{pre}}}{k_{\text{syn}}})};\\
    \end{aligned}
\end{equation}
Parameter $g^{(i,j)}_{\text{syn}}$ describes the synaptic weight, $g^{(i,j)}_{\text{syn}}=\eta+\nu^{(m,n)}_{Ca}$. The astrocyte $(m,n)$ modulates the synaptic currents of the neuron $(i,j)$. The variable $\nu_{Ca}$ introduces the astrocyte-induced modulation of synaptic strength and will be discussed below. Here, purely for illustration of astrocyte effects, we did not include intrinsic short-term synaptic plasticity in the model. The synaptic reversal potential for excitatory synapses is taken with $E_{\text{syn}}=0$. $V_{\text{pre}}$ denotes the membrane potential of the presynaptic neuron. For simplicity, we neglect the axonal and synaptic delays.

The architecture of synaptic connections between neurons is non-specific (e.g. random) with the following parameters. The number of output connections per each neuron is fixed at $N_{out}=40$. Each neuron innervates $N_{out}$ local postsynaptic targets, the distance  $R$ to which is determined according to the exponential distribution:
\begin{equation}
    \label{eq:Radius}
    \begin{aligned}
    f_R(R)=\begin{cases}
                \lambda\exp(-\lambda R), R \ge 0,  \\
                0, R < 0.\\
            \end{cases}
    \end{aligned}
\end{equation}

\begin{table}
      \centering
      \caption{Neural network parameters (\cite{Izhikevich2003,Kazantsev2011})}\label{tab_neuro}
      \begin{tabular}{ p{0.15\columnwidth}  p{0.65\columnwidth}  p{0.15\columnwidth} } 

          Parameter & Parameter description & Value  \\ 
          \hline 
         $W\times H$ & neural network grid size & $79\times 79$  \\ $a$ & time scale of the recovery variable  & 0.1  \\ 
          $b$ & sensitivity of the recovery variable
to the sub-threshold fluctuations of the membrane potential  & 0.2  \\
$c$ & after-spike reset value of the membrane potential  & -65 mV  \\
$d$ & after-spike reset value of the recovery variable  & 2  \\
 $\eta$ & synaptic weight without astrocytic influence  & 0.025  \\
 $E_{\text{syn}}$ & synaptic reversal potential for excitatory synapses ) & 0 mV  \\
 $k_{\text{syn}}$ & slope of the synaptic activation function  & 0.2 mV  \\
$N_{\text{out}}$ & number of output connections per each neuron & 40  \\
$\lambda$ & rate of the exponential distribution of synaptic connections distance & 5  \\
\hline           
      \end{tabular}
     \end{table}

\subsection{Astrocytic network}\label{sec_astro_model}
 
In the model we try to implement the biologically plausible organisation of astrocytic network and neuron-astrocyte interaction. The astrocytic network is configured in the form 
of a two-dimensional square lattice with dimension $M\times N$. Cortical astrocytes are coupled via CX43 gap junctions mostly permeable to inositol 1,4,5-trisphosphate (IP$_3$). Hence, in the model, we consider local diffusive coupling.  Besides, each astrocyte is interconnected with neuronal ensemble of $N_a$ neurons. It was experimentally shown that the sensory stimulation evokes fast intracellular Ca$^{2+}$ signals in fine processes of cortical astrocyte in response to local synaptic activity in neuronal circuit \citep{Stobart2018,Takata2011,Wang2006_2}. Multiple 
rapid, spatially restricted Ca$^{2+}$ events in the astrocytic process are induced by intense neuronal firing.  Local events are spatially and temporally integrated by the astrocytic cell, which results in a global, long lasting  Ca$^{2+}$ event. In turn, this event induces the release of gliotransmitters affecting synaptic transmission in the local territory of individual astrocytes  \citep{Henneberger2010,Bekar2008,Araque2014}. For simplicity, we did not model detailed process of spatial-temporal integration of the rapid Ca$^{2+}$ signals in the morphological structure of astrocyte modelled earlier by \citet{Gordleeva2019,Gordleeva2018} and \citet{Wu2018}. Here we employ a mean-field approach to describe the emergence of a global Ca$^{2+}$ signal and its impact on synchronisation of neuronal ensemble controlled by a certain astrocyte.  

As pyramidal neurons generate the spike, glutamate is released from the presynaptic terminal into the synaptic cleft (Fig.~\ref{fig_1pattern_trace}b,d). Amount of glutamate, $G$, that diffused from synaptic cleft and reached the astrocytic process can be described by the following equation \citep{Gordleeva2012,Pankratova2019}:
 \begin{equation}
\label{eq:Glu}
\begin{aligned}
\frac{dG^{(i,j)}}{dt} & = -\alpha_\text{glu} G^{(i,j)}+k_\text{glu}\Theta(V^{(i,j)}-30 mV),\\
\end{aligned}
\end{equation}
 here $\alpha_\text{glu}$ is the glutamate clearance constant, $k\text{glu}$ is the efficacy of the release, $\Theta$ denotes the Heaviside step function and $V^{(i,j)}$ is the membrane potential of the corresponding presynaptic neuron $(i,j)$. Binding of glutamate to metabotropic glutamate receptors (mGluR) on the astrocytic membrane, which is located close to the synapse, triggers the production of IP$_3$ in the astrocyte (Fig.~\ref{fig_1pattern_trace}e). We use the approaches from earlier studies to describe the dynamics of the intracellular concentration of IP$_3$ in the astrocyte \citep{Nadkarni2003,ULLAH2006}: 
 
\begin{equation}
\label{eq:ip3}
\begin{aligned}
\frac{dIP_3^{(m,n)}}{dt}  = \frac{IP_3^*-IP_3^{(m,n)}}{\tau_{IP3}}&+J_{\text{PLC$\delta$}}^{(m,n)}+J_{\text{glu}}^{(m,n)}+\text{diff}^{(m,n)}_{IP3},\\
\end{aligned}
\end{equation}
with $m = 1,\ldots, 26$, $n = 1,\ldots, 26$. Parameter $IP_3^*$ denotes the steady state concentration of the IP$_3$ and $J_{\text{PLC$\delta$}}$ describes the IP$_3$ production by phospholipase C$\delta$ (PLC$\delta$) (\cite{ULLAH2006}):
\begin{equation}
\label{eq:PLC}
\begin{aligned}
J_{\text{PLC}\delta} = \frac{v_4(Ca+(1-\alpha)k_4)}{Ca+k_4}
\end{aligned}
\end{equation}

The variable $J_{\text{glu}}$ describes the glutamate-induced production of the IP$_3$ in response to neuronal activity and is modelled as a rectangular-shaped pulse with amplitude $A_\text{glu}$ $\mu$M and duration $t_\text{glu}$ ms:
\begin{equation}
    \label{eq:Jglu}
    \begin{aligned}
    J_{\text{glu}} = \begin{cases}
                        A_\text{glu}, \qquad &\text{if} \quad t_0<t\le t_0+t_{\text{glu}},  \\
                        0, \qquad &\text{otherwise};\\
                     \end{cases}
    \end{aligned}
\end{equation}
here $t_0$ denotes the periods when the total level of glutamate in all synapses associated with this astrocyte reaches a threshold:
\begin{equation}
\label{eq:Nact}
\begin{aligned}
\left(\frac{1}{N_a}\sum_{(i,j)\in{N_a}}[G^{(i,j)}>G_\text{thr}]\right)>F_{act}, 
\end{aligned}
\end{equation}
here we use the parameter $G_\text{thr}=0.7$. $[]$ denotes the Iverson bracket. $F_\text{act}$ is the fraction of synchronously spiking neurons of the neuronal ensemble corresponding to the astrocyte. For the emergence of the calcium elevation $F_\text{act}= 0.5$ is required. In other words, the production term, $J_{\text{glu}}$, is activated when correlated activity in the neuronal ensemble reaches a certain level of coherence.

Increase of IP$_3$ concentration in the astrocyte induces the release of Ca$^{2+}$ from internal stores, mostly from the endoplasmic reticulum (ER), to cytosol. For simplified description of the biophysical mechanism underlying the calcium dynamics in astrocytes, we use the Ullah model \citep{ULLAH2006}. Changes of the intracellular Ca$^{2+}$ concentration, $Ca$, are described by the following equations:
\begin{equation}
\label{eq:astro_main}
\begin{aligned}
&\frac{dCa^{(m,n)}}{dt}  = J_{\text{ER}}^{(m,n)}-J_{\text{pump}}^{(m,n)}+J_{\text{leak}}^{(m,n)}+J_{\text{in}}^{(m,n)}-J_{\text{out}}^{(m,n)}+\text{diff}^{(m,n)}_{Ca};\\
&\frac{dh^{(m,n)}}{dt}  = a_2 \left(d_2\frac{IP_3^{(m,n)}+d_1}{IP_3^{(m,n)}+d_3}(1-h^{m,n})-Ca^{(m,n)}h^{(m,n)} \right);
\end{aligned}
\end{equation}
where $h$ is the fraction of the activated IP$_3$ receptors (IP$_3$Rs) on the ER surface. Flux $J_{\text{ER}}$ is Ca$^{2+}$ flux from the ER to the cytosol through IP$_3$Rs, $J_{\text{pump}}$ is the flux pumped Ca$^{2+}$ back into ER via the sarco/ER Ca$^{2+}$-ATPase (SERCA), $J_{\text{leak}}$ is the leakage flux from the ER to the cytosol. Fluxes $J_{\text{in}}$ and $J_{\text{out}}$ describe calcium exchange with extracellular space. The fluxes are expressed as follows:
\begin{equation}
\label{eq:astro_currents}
\begin{aligned}
&J_{\text{ER}} = c_1v_1Ca^3h^3IP_3^3\frac{(c_0/c_1-(1+1/c_1)Ca)}{((IP_3+d_1)(IP_3+d_5))^3};\\
&J_{\text{pump}} = \frac{v_3Ca^2}{k_3^2+Ca^2};\\
&J_{\text{leak}} = c_1v_2(c_0/c_1-(1+1/c_1)Ca);\\
&J_{\text{in}} = \frac{v_6IP_3^2}{k_2^2+IP_3^2};\\
&J_{\text{out}} = k_1Ca;\\
\end{aligned}
\end{equation}
Biophysical meaning of all parameters in Eqs. \eqref{eq:ip3}, \eqref{eq:PLC},\eqref{eq:astro_main},\eqref{eq:astro_currents} and their values determined experimentally can be found in \citep{Li1994,ULLAH2006} and in Table 2.

Cortical astrocytes are coupled by CX43 gap junctions \citep{Nimmerjahn2004}. Thus, diffusion of active chemicals becomes possible between the neighbouring astrocytes. Currents $\text{diff}_{Ca}$ and $\text{diff}_{IP_3}$ describe the diffusion of Ca$^{2+}$ ions and IP$_3$ molecules via gap junctions between the astrocytes in the network and can be expressed as follows:
\begin{equation}
\label{eq:astro_dif}
\begin{aligned}
&\text{diff}^{(m,n)}_{Ca} = d_{Ca}(\Delta Ca)^{(m,n)};\\
&\text{diff}^{(m,n)}_{IP3} = d_{IP3}(\Delta IP_3)^{(m,n)};\\
\end{aligned}
\end{equation}
where parameters $d_{Ca}$ and $d_{IP_3}$ describe the Ca$^{2+}$ and IP$_3$ diffusion rates, respectively. Following experimental data, we assume that CX43 is less permeable to $Ca^{2+}$ than to  $d_{IP_3}$.  $(\Delta Ca)^{(m,n)}$ and $(\Delta IP_3)^{(m,n)}$ are the discrete Laplace operators:  
\begin{equation}
\label{eq:astro_Laplace}
\begin{aligned}
(\Delta Ca)^{(m,n)}&=(Ca^{(m+1,n)}+Ca^{(m-1,n)}+Ca^{(m,n+1)}+Ca^{(m,n-1)}-4Ca^{(m,n)}).\\
\end{aligned}
\end{equation}
Equations \eqref{eq:ip3}-\eqref{eq:Jglu},\eqref{eq:astro_main}-\eqref{eq:astro_dif} predict that the  synchronized activity in the neuronal ensemble triggers the astrocytic Ca$^{2+}$ signals, and in the absence of neuronal stimulus in the astrocytic network steady state Ca$^{2+}$ concentration is maintained. 

Next, we account for the effect of enhancement of excitatory postsynaptic currents ($ePSC$) generation through modulation of postsynaptic NMDARs by D-serine released from the astrocyte \citep{Bergersen2011,Henneberger2010}. In the model, we propose that global events of Ca$^{2+}$ elevation in astrocyte result in D-serine release, which can modulate the synaptic strength of all synapses corresponding to the morphological territory of this astrocyte. For simplicity, the relationship between the astrocyte Ca$^{2+}$ concentration  and synaptic weight of the affected synapses $g_{\text{syn}}$, is described as follows:
\begin{equation}
\label{eq:astro_neuro}
\begin{aligned}
g_{\text{syn}}&=\eta+\nu_{Ca},\\
\nu_{Ca} &= \nu_{Ca}^*\Theta(Ca^{(m,n)}-Ca_\text{thr} ) 
\end{aligned}
\end{equation}    
where the parameter $\nu_{Ca}^*$ denotes the strength of the astrocyte-induced modulation of the synaptic weight, $\Theta(x)$ is the Heaviside step-function. The feedback from the astrocytes to the neurons is activated when the astrocytic Ca$^{2+}$ concentration is larger than $Ca_\text{thr}$ and the fraction of synchronously spiking neurons of neuronal ensemble corresponding to the astrocyte $F_\text{astro}$ during the time period of $\tau_\text{syn} = 10$ ms. The duration of the feedback is fixed and is equal to $\tau_\text{astro}= 250$ ms.  

\begin{table}
      \centering
      \caption{Astrocytic network parameters \citep{ULLAH2006}}\label{tab_astro}
      \begin{tabular}{ p{0.15\columnwidth}  p{0.6\columnwidth}  p{0.2\columnwidth} } 

          Parameter & Parameter description & Value  \\ 
          \hline 
         $M\times N$ & astrocytic network grid size & $26\times 26$  \\ $c_0$  & total Ca$^{2+}$ in terms of cytosolic vol  & 2.0 $\mu$M  \\ 
          $c_1$ & 	(ER vol)/(cytosolic vol)  & 0.185  \\
$v_1$ & max Ca$^{2+}$ channel flux & 6 s$^{-1}$  \\
$v_2$ & Ca$^{2+}$ leak flux constant  & 0.11 s$^{-1}$  \\
 $v_3$ & max Ca$^{2+}$ uptake & 2.2 $\mu$M s$^{-1}$  \\
 $v_6$ & maximal rate of activation dependent calcium influx & 0.2 $\mu$M s$^{-1}$  \\
$k_1$ & rate constant of calcium extrusion & 0.5 s$^{-1}$ \\
$k_2$ & half-saturation constant for agonist-dependent calcium entry & 1 $\mu$M \\
$k_3$ & activation constant for ATP-Ca$^{2+}$ pump & 0.1 $\mu$M \\
$d_1$ & dissociation constant for IP$_3$ & 0.13 $\mu$M \\
$d_2$ & dissociation constant for Ca$^{2+}$ inhibition & 1.049 $\mu$M \\
$d_3$ & receptor dissociation constant for IP$_3$ & 943.4 nM \\
$d_5$ & Ca$^{2+}$ activation constant & 82 nM	 \\
$\alpha$ & & 0.8 \\
$v_4$ & max rate of IP$_3$ production & 0.3 $\mu$M s$^{-1}$ \\
1/$\tau_r$ & rate constant for loss of IP$_3$ & 0.14 s$^{-1}$ \\
${IP_3}^*$ & steady state concentration of IP$_3$ & 0.16 $\mu$M \\
$k_4$ & dissociation constant for Ca$^{2+}$ stimulation of IP$_3$ production & 1.1 $\mu$M \\
$d_{Ca}$ & Ca$^{2+}$ diffusion rate & 0.05 s$^{-1}$ \\
$d_{IP_3}$ & IP$_3$ diffusion rate & 0.1 s$^{-1}$ \\
\hline           
      \end{tabular}
     \end{table}

\begin{table}
      \centering
      \caption{Neuron-astrocytic interaction parameters \citep{Gordleeva2012}}\label{tab_astro}
      \begin{tabular}{ p{0.1\columnwidth}  p{0.6\columnwidth}  p{0.2\columnwidth} }
          Parameter & Parameter description & Value \\ 
          \hline 
         $N_a$ & number of neurons interacting with one astrocyte & 16, $4\times 4$  \\ 
         $\alpha_\text{glu}$  & glutamate clearance constant & 10 s$^{-1}$  \\ 
          $k_\text{glu}$ & efficacy of the glutamate release & 600 $\mu$M s$^{-1}$ \\
          $A_{glu}$ & rate of IP$_3$ production through glutamate & 5 $\mu$M s$^{-1}$ \\
          $t_\text{glu}$ & duration of IP$_3$ production through glutamate & 60 ms \\
$G_\text{thr}$ & threshold concentration of glutamate for IP$_3$ production & 0.1  \\
$F_\text{act}$ & fraction of synchronously spiking neurons required for the emergence of Ca$^{2+}$ elevation & 0.5  \\
$F_\text{astro}$ & fraction of synchronously spiking neurons required for the emergence of astrocytic modulation of synaptic transmission & 0.375  \\
 $\nu_{Ca}^*$ & strength of the astrocyte-induced modulation of synaptic weight & 0.5  \\
 $Ca_\text{thr}$ & threshold concentration of Ca$^{2+}$ for the astrocytic modulation of synapse & 0.15 $\mu$M \\
 $\tau_\text{astro}$ & duration of the astrocyte-induced modulation of synapse & 250 ms  \\
\hline           
      \end{tabular}
     \end{table}

\subsection{Stimulation protocol}\label{sec_stimul}
The term $I_{app}$ in Eq. \eqref{eq:Izh_main} represents specific and non-specific external inputs. A non-specific noisy input simulates the input signals from networks of other brain areas and is applied continuously to all neurons in the form of independent Poisson pulse trains of a certain rate, $f_\text{bg}$, with amplitudes randomly and uniformly distributed in the interval $[-10, 10]$ $\mu$A. This input evokes background network state with low-rate, spontaneous spiking. 

Specific input contains training samples in the form binary spatial patterns. The patterns represent different spatial distributions relative to background state with non-specific input only. Average size of sample is 1078 neurons ($18$\% of network) stimulated by the specific input, with average $35.2$\% overlapping in the population. For visual representation of samples, we take binary images of numerals (0,1,2,3,4,..) with size $W\times H$ pixels, where each pixel corresponds to a neuron in the neuronal layer. Neurons corresponding to the shape of the numerals receive rectangular excitatory pulse with length $t_\text{stim}$ and amplitude $A_\text{stim}$. The shape of each sample was spatially distorted by 5\% random noise such as “salt and pepper noise”. Then a transient inputs were applied to simulate the nonmatching test items and the cue (length $t_\text{test}$, and amplitude $A_\text{test}$). In the cued recall for simulating the loss in saliency, we applied shorter input with lower amplitude and higher level (20\%) of random noise. 

\begin{table}
      \centering
      \caption{Stimulation protocol and recall testing parameters}\label{tab_astro}
      \begin{tabular}{ p{0.1\columnwidth}  p{0.6\columnwidth}  p{0.2\columnwidth} } 
          Parameter & Parameter description & Value \\ 
          \hline 
         $f_{bg}$ & background activity rate & 1.5 Hz  \\ 
         $A_\text{stim}$  & stimulation amplitude & 10 $\mu$A  \\ 
          $t_\text{stim}$ & stimulation duration & 200 ms \\
           & noise level in sample & 5$\%$  \\
           $A_\text{test}$  & cue stimulation amplitude & 8 $\mu$A  \\ 
          $t_\text{test}$ & cue stimulation length & 150 ms \\
           & noise level in cue & 20$\%$  \\
\hline           
      \end{tabular}
     \end{table}

\subsection{Memory performance metrics}\label{sec_analysis}
To measure memory performance of the system we count the correlation of recalled pattern with ideal item in the following way:
\begin{equation}
\label{eq:accuracy}
\begin{aligned}
&M_{ij}(t)=I\left[\left(\sum_{k=t-w}^t I[V_{ij}(k)>thr]\right)>0 \right],\\
&C(t)=\frac{1}{2}\left(\frac{1}{|P|}\sum_{(i,j)\in P}M_{ij}(t)+\frac{1}{W\cdot H-|P|}\sum_{(i,j)\notin P} (1-M_{ij}(t))\right),\\
&C_P=\frac{1}{|T_P|}\max_{t\in T_P} C(t);
\end{aligned}
\end{equation} 
here $w=10\text{ frames}=1$ ms, $P$ - a set of pixels belonging ideal pattern, $W,H$ - network dimension, $thr$ - spike threshold, $I$ - indicator function, $T_P$ - a set of frames in the tracking range of pattern $P$. In a sense, this correlation metric can be associated with $1-d$ averaged between pattern and background, where $d$ is the Hamming distance. 

\section{Results}
Let us show how the neuron-astrocyte network model exhibits memory formation. First, we will consider a simple single-item memory task illustrating information loading, storage, and retrieval. Next, we will demonstrate how the network can be successfully trained to memorize and recall several patterns with significant overlaps. Finally, we will analyse model performance metrics, capacity, and characteristic of pattern remembering on different parameters.

\subsection{Single-item WM}
\begin{figure*}
\includegraphics[width=0.9\textwidth]{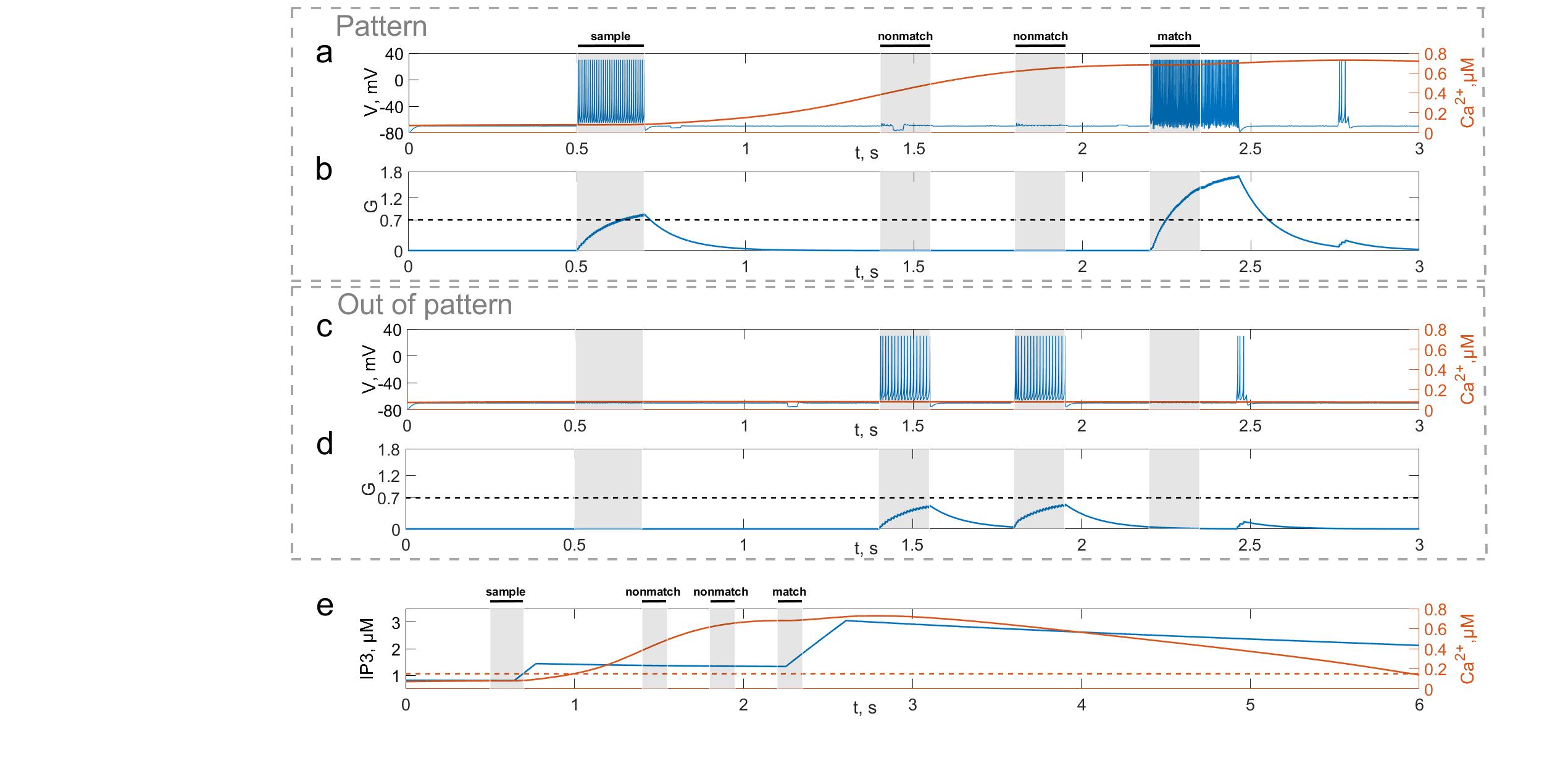}
\caption{Model of neuron-astrocyte interaction. (a) Spike train and (b) concentration of  neurotransmitter, $G(t)$, of stimulus-specific neuron. (c,d) Same as in (a,b) but for an unspecific neuron. (e) Intracellular concentration of Ca$^{2+}$ and IP$_3$ in stimulus-specific astrocyte. Black bars at the top indicate periods when each of the stimuli (training stimulus - sample, nonmatching test items - nonmatch, test cue - match) was presented. In response to presynaptic spike train (a,c), the neurotransmitter, glutamate, $G$ releases (b,d) into extracellular space and the concentration of IP$_3$ increases in astrocyte (e, blue line) inducing the elevation of intracellular Ca$^{2+}$ (e, red line).} \label{fig_1pattern_trace}
\end{figure*}

\begin{figure*}
\includegraphics[width=0.9\textwidth]{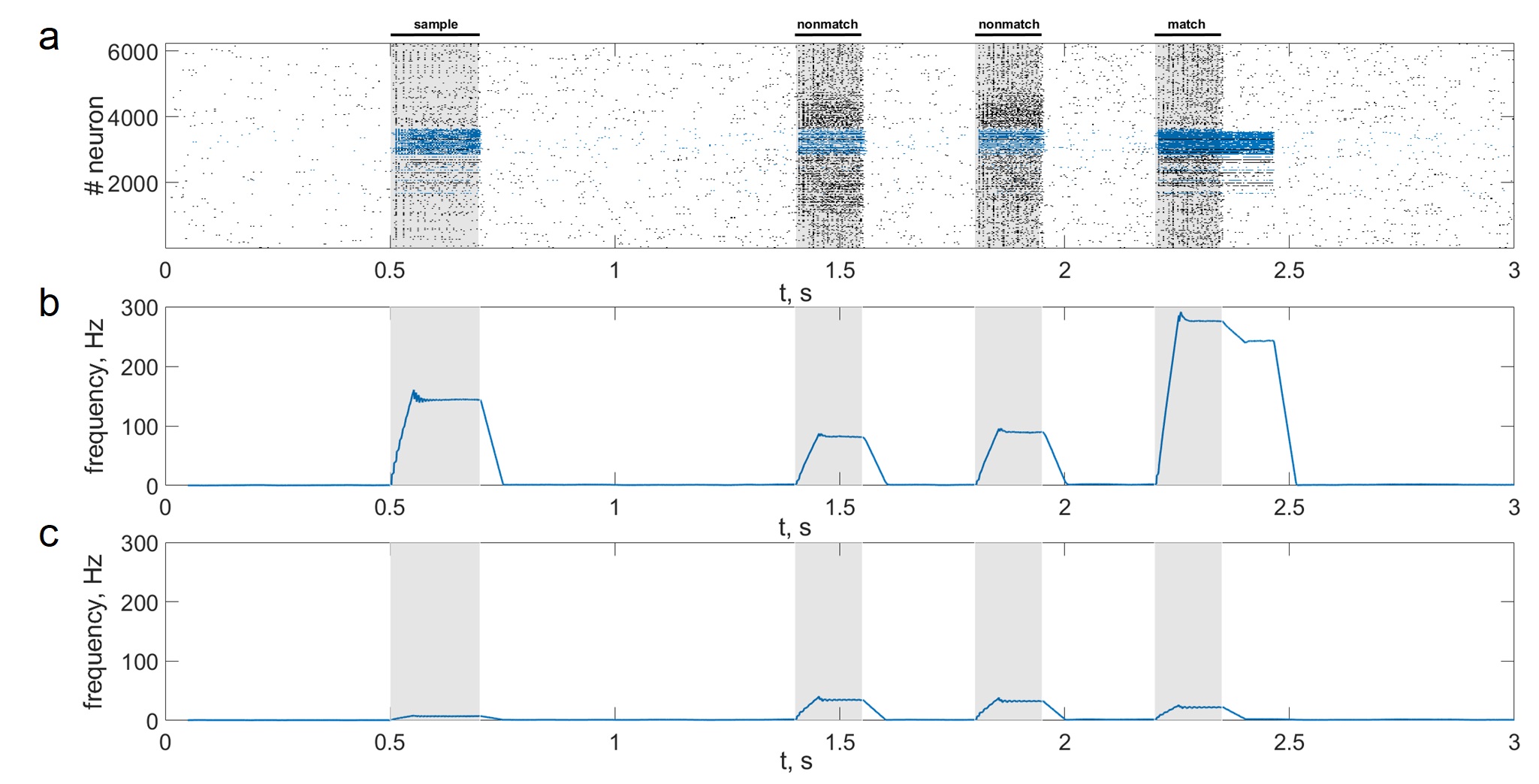}
\caption{Delayed matching to sample WM task. (a) One trial of the task simulated in the network. Spike raster of neuronal network showing sample-selective delay activity. Neurons belonging to stimulus-specific population are indicated by red color. Black bars indicate periods when each of the stimuli was presented. (b, c) The averaged firing rate of the stimulus-specific and unspecific neurons over time, respectively. (20 ms bins) }\label{fig_1pattern_rastr}
\end{figure*}

First, we will test the neuron-astrocyte network in the most common experimental paradigm of WM studies - the DMS task. This task requires a single item to be held in memory during a brief delay period. Before specific stimulation, neural network demonstrates irregular, low-rate background activity (see activity beginning in Fig \ref{fig_1pattern_rastr}). At the 500 ms mark, we load an item by applying transient external input to the corresponding neuronal population for 200 ms (Figs. \ref{fig_1pattern_trace}a, \ref{fig_1pattern_rastr}). During training, each astrocyte tracks the activity of neuronal subnetwork associated with it. As soon as the extracellular concentration of glutamate (Figs. \ref{fig_1pattern_trace}b,d) and correlated firing in neurons achieve a certain level, which satisfies the condition Eq. (\ref{eq:Nact}), Ca$^{2+}$ concentration in matching astrocytes elevates (Fig. \ref{fig_1pattern_trace}e). In accordance with the experimental data \citep{Bindocci2017}, we tuned the model parameters in such way that the onset of the calcium elevation in astrocytes induced by synchronous neuronal discharge has a delay of  $\le$ 2 sec. Following the increased firing in the stimulus-specific part of the neuronal network upon reaching a threshold of 0.15 $\mu$M the astrocyte releases gliotransmitters modulating the synaptic strengths in corresponding locations  (Fig. \ref{fig_1pattern_trace}).  
The calcium pulse in astrocyte lasts for several seconds. Its duration determines the length of the delay interval in the DMS task, during which the item is maintained in the memory. After the training stimulus ends, we test maintenance of the single-item memory by applying two nonmatching items and cue item with $t_\text{test}$ durations and 250 ms inter-item interval (Figs. \ref{fig_1pattern_trace}, \ref{fig_1pattern_rastr}). Because the astrocytic feedback also depends on the activity of neuronal subnetwork, the model responds differently to the applied items. A short presentation of the cue to the neural network evokes the astrocytic-induced increase in the synaptic strength between stimulus-specific neurons and results in a local spatial synchronization in the whole stimulus-specific neuronal population  (Fig. \ref{fig_1pattern_trace}a with comparison to Fig. \ref{fig_1pattern_trace}c). Similar to experimental data \citep{Miller1996}, delay activity in our model is sample-selective. We observe that the pattern-specific firing rate in the neuronal network increases and is equal to 270 Hz in comparison with response to non-specific stimulus (80 Hz) (Fig. \ref{fig_1pattern_rastr},b). Such a high frequency is determined by the choice of fast-spiking neuron model \citep{Izhikevich2003}. The firing rates in simulations with regular spiking neuron model \citep{Izhikevich2003} are almost 10 times lower: 30 Hz for stimulus-specific and 4.5 Hz for non-specific stimulus. The elevation of the frequency in the stimulus-specific neuronal population can continue after the end of the cue, which is determined by duration of the astrocyte-induced enhancement of the synaptic weight. 

\begin{figure*}
\includegraphics[width=0.9\textwidth]{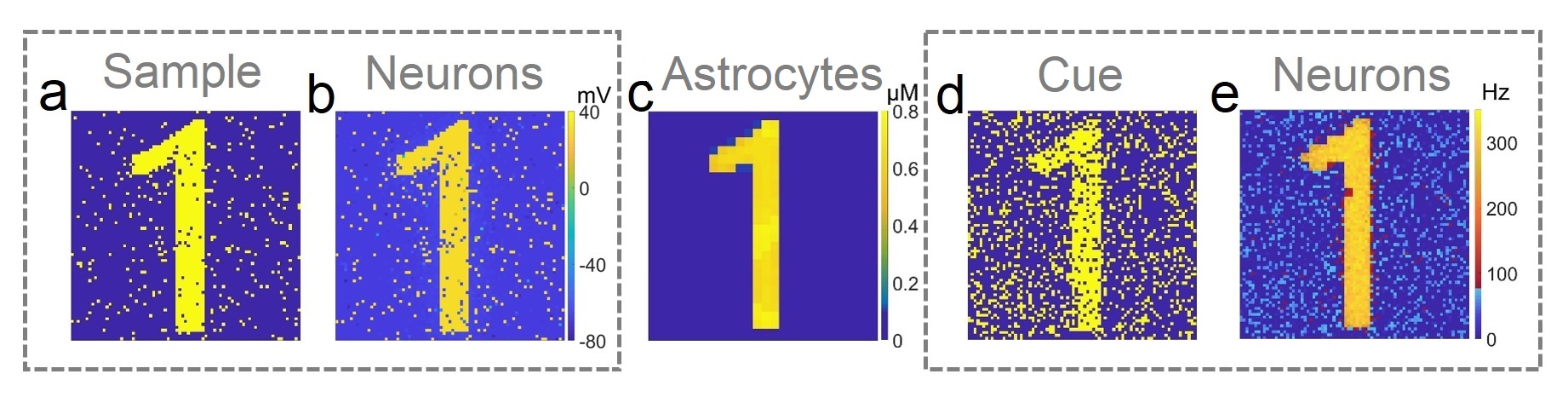}
\caption{Snapshots of the training (a,b,c) and testing (d,e) neuron-astrocyte network in the single-item working memory task. (a) Training sample. (b) Response of the neuronal network to the sample. The values of the membrane potentials are shown. (c) Intracellular Ca$^{2+}$ concentrations in astrocytic layer. (d) Testing item with 20\%  salt and pepper noise. (e) Cued recall in the neuronal network. The firing rate averaged on the test time interval for each neuron is shown. }\label{fig_1pattern_snapshots}
\end{figure*}

For a visual representation of memory formation, we follow space-time distribution of sample-selective delay activity. Figure (\ref{fig_1pattern_snapshots}) illustrates the spatial distribution of activity in neuronal and astrocytic layers at the different moments of training and cued recall for the same single-item memory task as presented in the Figure  (\ref{fig_1pattern_rastr}). Training induces the emergence of synchronized calcium activity of spatially clustered astrocytes (Fig. \ref{fig_1pattern_snapshots}c). Note that locally synchronised astrocytes have been found in the neocortex and hippocampus in situ and in vivo  \citep{Sasaki2011,Takata2008}. Such calcium activity correlated in time and space can lead to the spatial-temporal synchronization in the neuronal network \citep{Araque2014}. This mechanism of neuron-astrocyte network interaction underlies the sample-selectivity and pattern retrieval in the model. Note that 20 $\%$ noisy cue item (Fig. \ref{fig_1pattern_snapshots}d) can be identified and cleared from noise by the neuronal network due to the astrocyte-induced feedback (Fig. \ref{fig_1pattern_snapshots}e). In other words, the spiking neuronal network accompanied by astrocytes can filter a cue pattern distorted by noise. Video of the single-item memory encoding and cued recall in the neuron-astrocyte network can be found in the supplementary material.

\subsection{Multi-item WM}
\begin{figure*}
\includegraphics[width=0.9\textwidth]{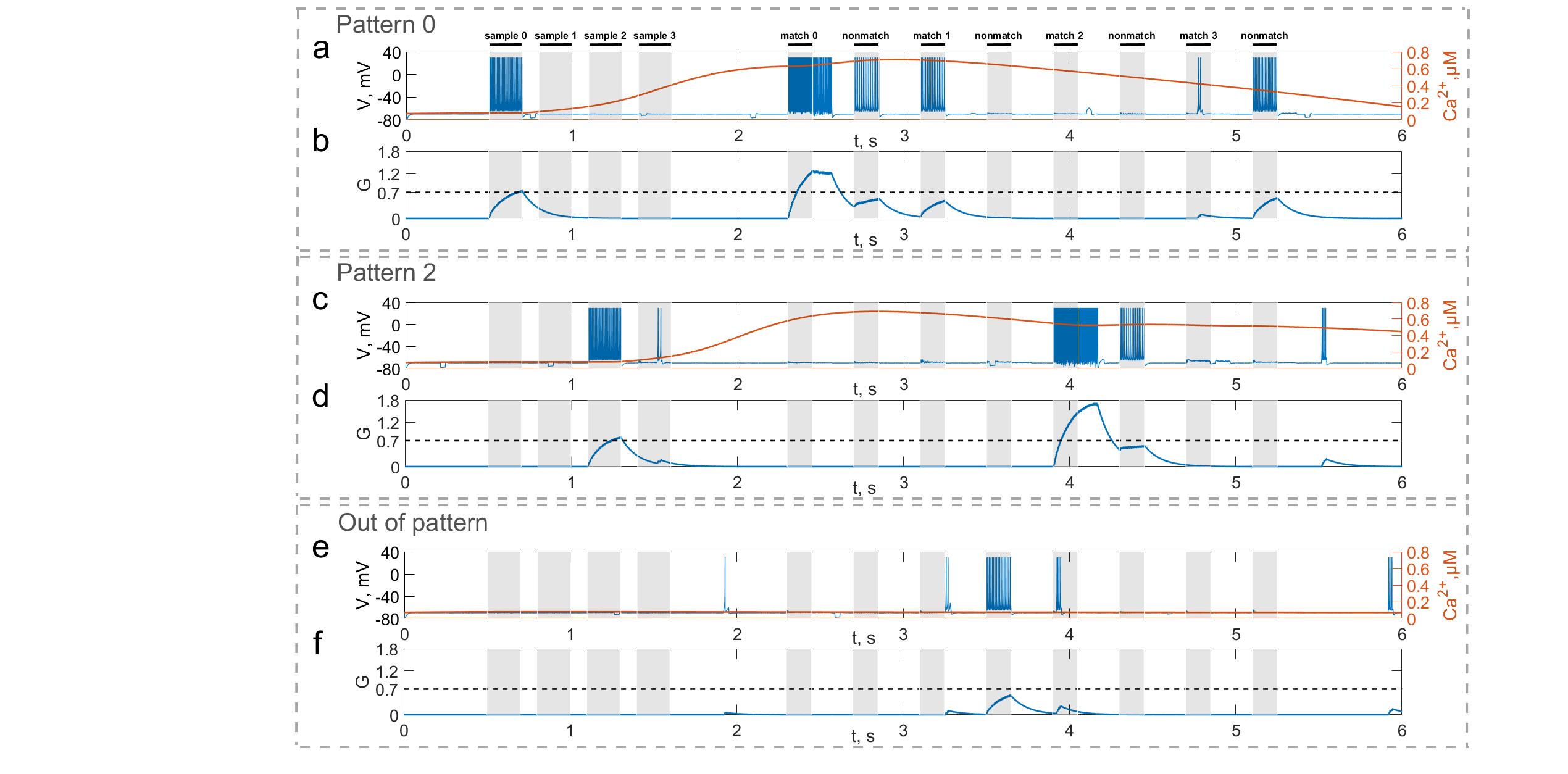}
\caption{Neuron-astrocyte network simulation with four loaded memory items. (a,c,e) Spike train and (b,d,f) concentration of neurotransmitter, $G(t)$, of three neurons belonging to different stimulus-specific populations. (a,b) Stimulus-specific neuron to sample 0. (c,d) Stimulus-specific neuron to sample 2. (e,f) Neuron unspecific to all samples. Black bars at the top indicate periods when each of the stimuli (training stimulus - sample, nonmatching test items - nonmatch, test cue - match) was presented.} \label{fig_5pattern_trace}
\end{figure*}

\begin{figure*}
\includegraphics[width=1.0\textwidth]{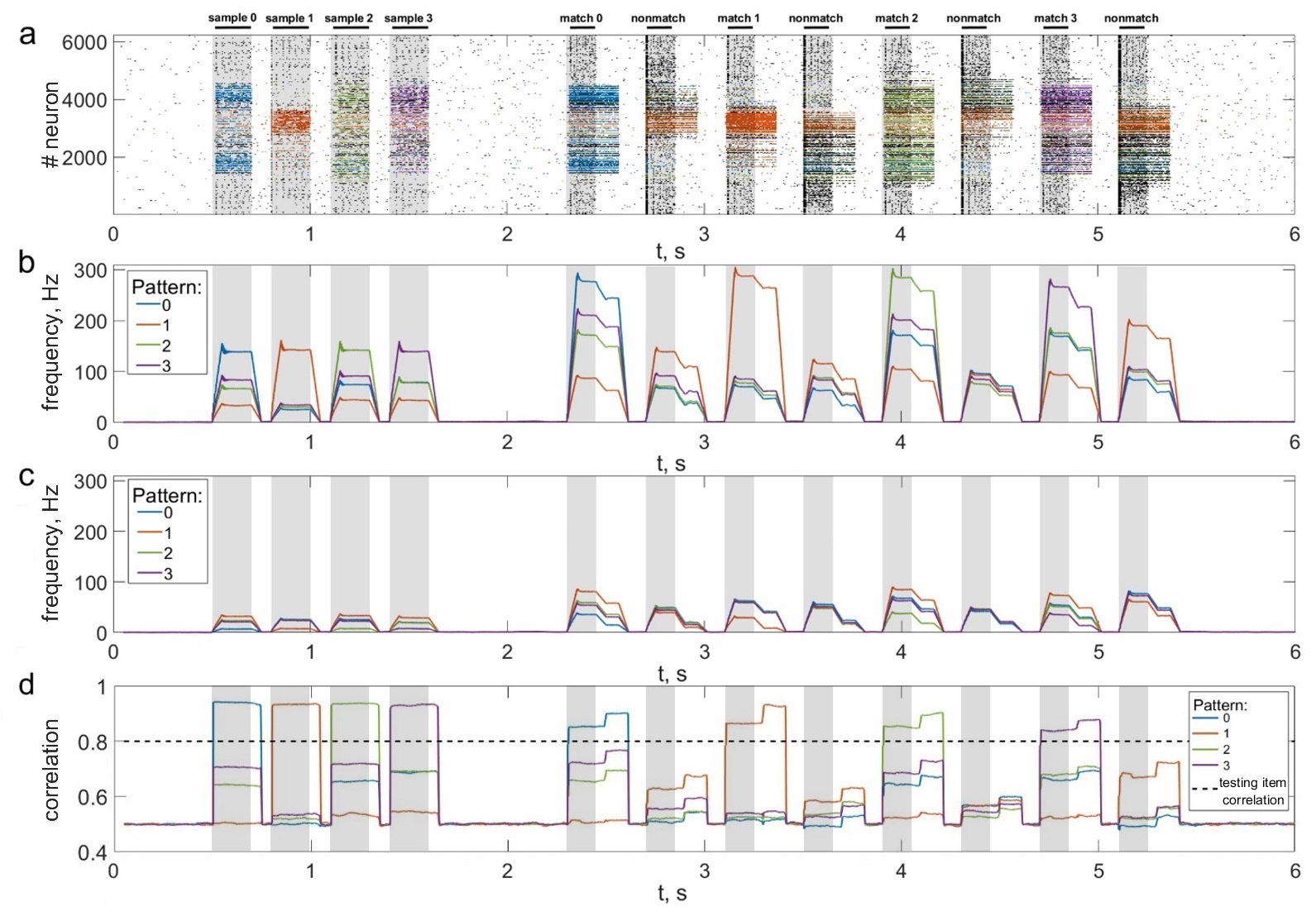}
\caption{Multi-item WM in neuron-astrocyte network. (a) Spike raster of neuronal network with 4 training patterns. Neurons are coloured according to their pattern selectivity. Pattern overlapping in neuronal populations is $35.2$\% on average for 4 patterns. Black bars indicate periods when each of the stimuli was presented. (b, c) The averaged firing rate of the stimulus-specific and unspecific neurons over time, respectively. (20 ms bins) (d) Correlation of  filtered items. The different colours correspond to the correlations with different ideal samples. The dotted line shows the correlation of the testing item (for 20\% noise level in test). }\label{fig_5pattern_rastr}
\end{figure*}

\begin{figure*}
\includegraphics[width=0.9\textwidth]{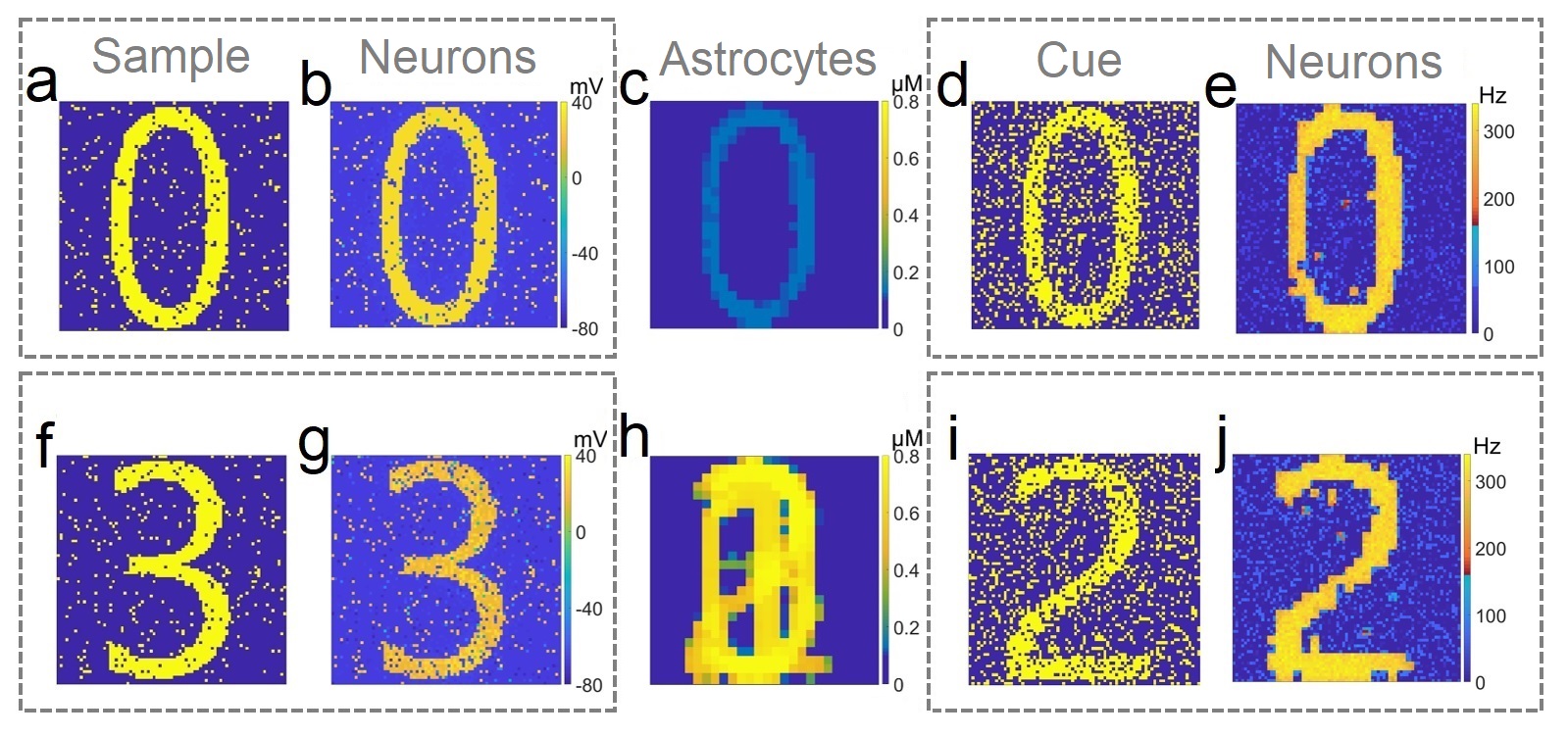}
\caption{Snapshots of the training (a, b, c, f, g, h) and testing (d, e, i, j) neuron-astrocyte network in the multi-item working memory task. We used the following training set consisting of 4 samples: 0,1,2,3. (a,f) The example of first and last training samples, respectively. (b,g) The response of the neuronal network to samples. The values of the membrane potentials are shown. (c,h) The intracellular Ca$^{2+}$ concentrations in astrocytic layer. (d,i) The testing items with 20\% of salt and pepper noise. (e,j) The cued recalls in the neuronal network. The firing rate averaged on the test time interval for each neuron is shown. }\label{fig_4pattern_snapshots}
\end{figure*}

Next, we consider the multi-item WM formation. In this case, we loaded four items, images of numerals 0,1,2,3 in the following way. The images were loaded consequently by applying external inputs of $t_\text{stim}$ durations with 100 ms inter-item intervals (see Figs. \ref{fig_5pattern_trace}, \ref{fig_5pattern_rastr}a). Due to the coincidence of different stimulus-specific neuronal populations in space, the spatial calcium patterns in astrocytic layers for different items overlap significantly (Fig. \ref{fig_4pattern_snapshots}h). After a 700 ms training stimulus was applied, we tested the maintenance of the memory by applying matching and nonmatching items of $t_\text{test}$ durations and 250 ms inter-item intervals (Figs. \ref{fig_5pattern_trace}, \ref{fig_5pattern_rastr}a). The images were distorted by 20\% noise. The astrocyte-mediated feedback modulating coherent neuronal activity provided the selectivity of the model response. The system remembered the correct image. Thus, we observed that all items were successfully filtered only in the cued recall. Needless to say that firing rate increases significantly in the cued recall due to the selective increase of synaptic strengths  (Figs. \ref{fig_5pattern_rastr}b,c). To evaluate the performance of the neuron-astrocyte WM, we used as metric the correlation between recalled item and the ideal item during the multi-item WM task (see section Memory performance metrics) (Fig. \ref{fig_5pattern_rastr}d). It is important to note, that during multi-item remembering spurious correlations never dominate in that sense accuracy of our system is always equal to 100\%. There was an increase in correlation with the target image and no attraction to the wrong image or chimeras. Maximal correlation reached 95\% in training and 93\% in testing sets on average for 4 samples. Video of the multi-item WM in the neuron-astrocyte network can be found in the supplementary material 2.

\begin{figure}
	\centering
	\includegraphics[width=0.9\columnwidth]{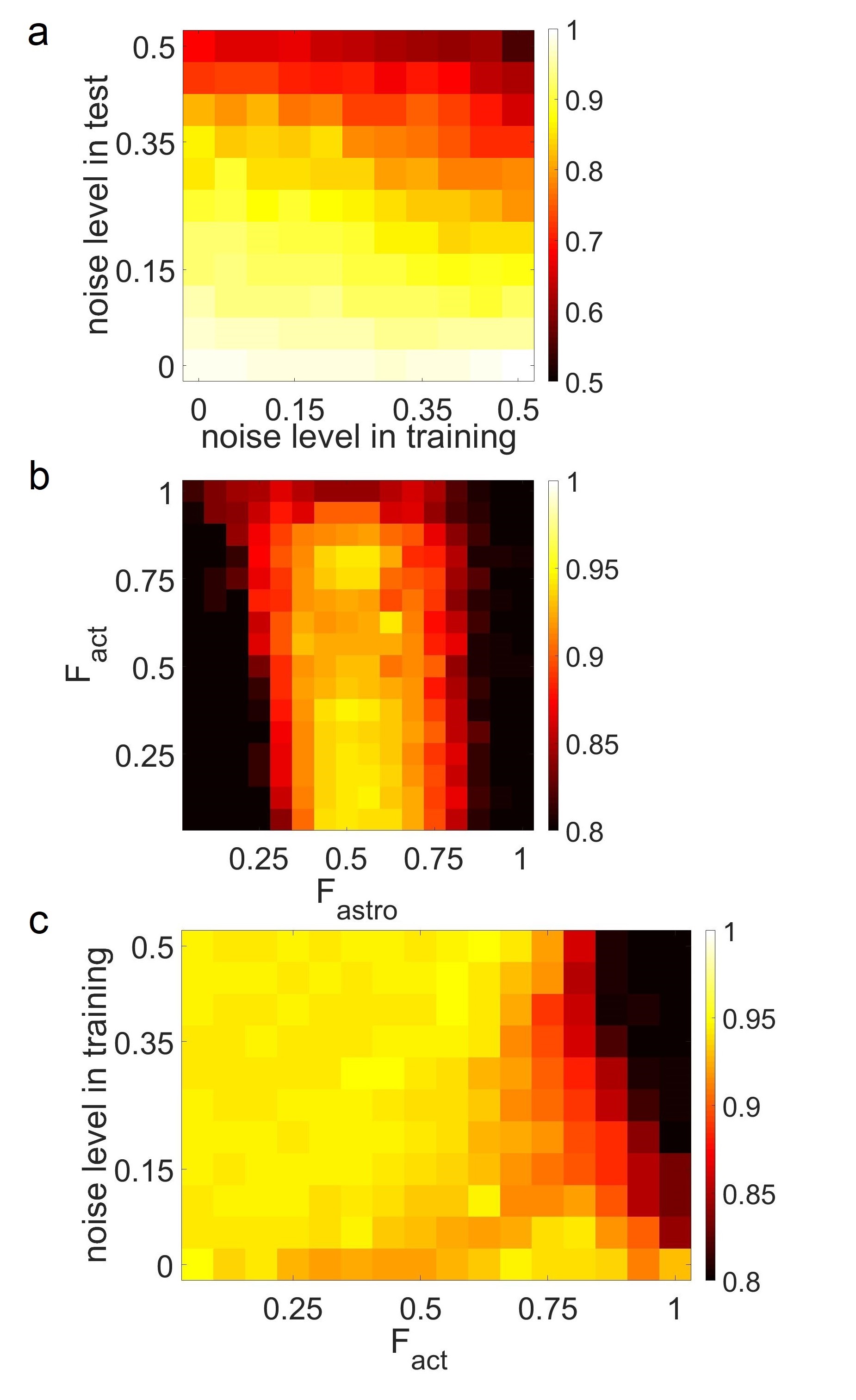}
	\caption{Correlation between recalled pattern (see section \ref{sec_analysis}) and ideal item in a multi-item WM task performed by the neuron-astrocyte network. The correlation averaged over 4 patterns is shown. (a) Noise-resistance ability of the model. Dependence of correlation on noise level in training and test. The correlation difference between cued recall pattern and noisy input is shown. (b),(c) The influence of the neuron-astrocytic interaction structure. (b) Dependence of correlation on the number of spiking neurons required for the calcium elevation in the astrocyte, $F_{act}$, and on the number of spiking neurons required for the emergence of astrocyte-induced enhancement synaptic transmission, $F_{astro}$. (c) Dependence of correlation on noise level in training samples and on the parameter, $F_{act}$. $F_{astro}=0.5$. For (b,c) noise level in cue is 20\%.}\label{fig_accuracy}
\end{figure}

\begin{figure}
	\centering
	\includegraphics[width=1\columnwidth]{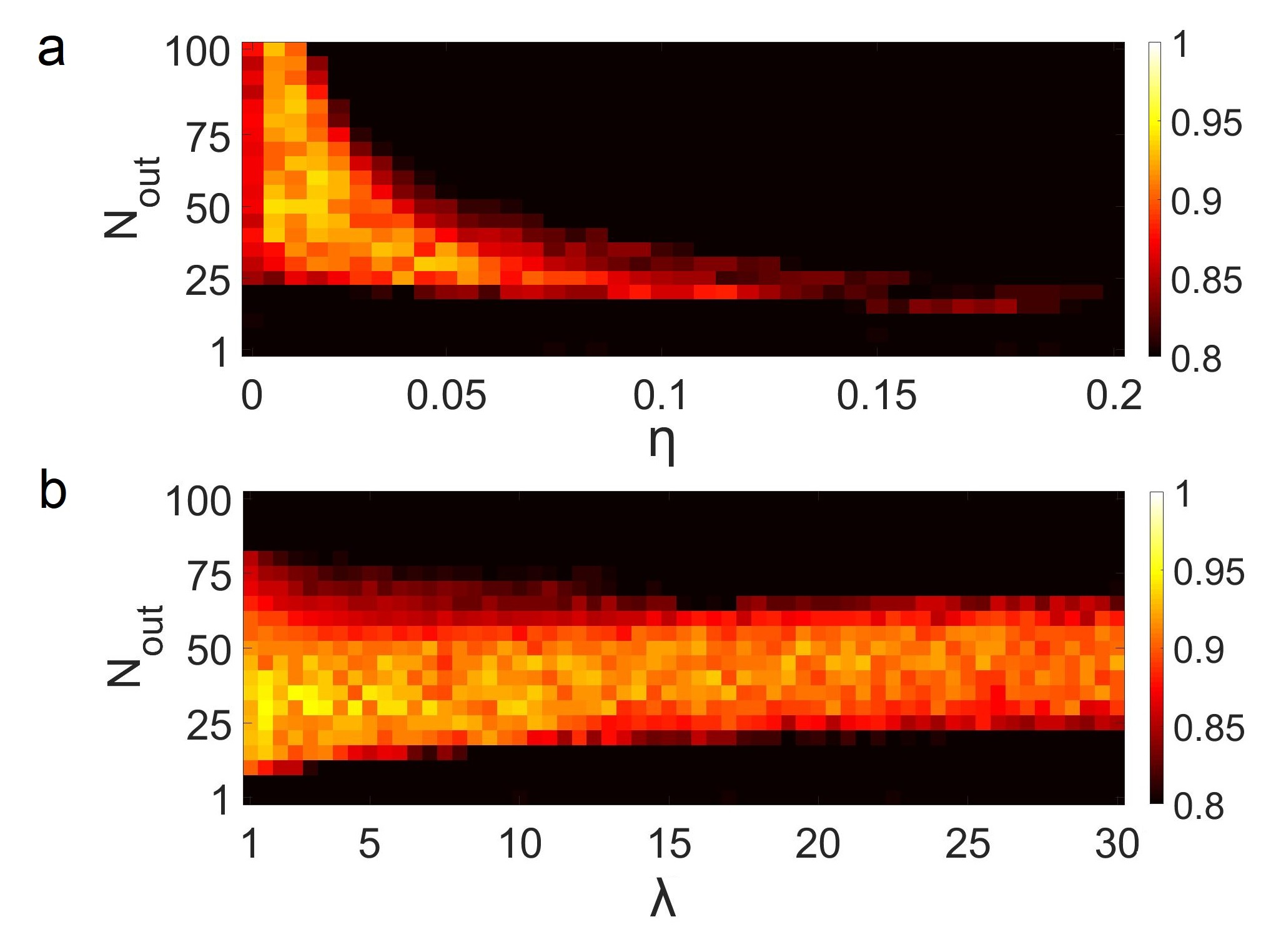}
	\caption{Influence of synaptic connectivities architecture in the neural network on the correlation of recalled pattern in multi-item WM task performed by the neuron-astrocyte network. The correlation averaged over 4 patterns is shown. (a) Dependence of correlation on the number of output synaptic connections for each neuron, $N_\text{out}$, and synaptic weight, $\eta$. (b) Dependence of correlation on the number of output synaptic connections for each neuron, $N_\text{out}$, and synaptic connection distribution parameter, $\lambda$ Eq.(\ref{eq:Radius}). }\label{fig_accuracy_2}
\end{figure}

To characterise the quality of memory formation in the model, we examined the dependencies of correlation of retrieval pattern in cued recall on variable parameters of input patterns, astrocytic, synaptic, and network structure (Figs. \ref{fig_accuracy},\ref{fig_accuracy_2}). First, we investigated its dependence on noise parameters. The dependence of correlation of recalled pattern on the noise level in training and test experiments is shown in Figure (\ref{fig_accuracy}a). Specifically, the correlation difference between the recalled pattern and noisy input is presented. In other words, the model can improve test images depending on noise in training and testing. 
Training the network with samples with a low noise level (up to 25\%) provides high correlation. The elevation of the noise level in training sample induces a random activity pattern in astrocytic network, which in turn leads to noisy recall. 

The morpho-functional structure of connections between neurons and astrocytes can affect pattern retrieval in the model (Figs. \ref{fig_accuracy}b,c). The key parameters determining this structure are the fraction of synchronously spiking neurons of neuronal ensemble corresponding to the astrocyte required for the emergence of the calcium elevation in the astrocyte, $F_{act}$, and the fraction of synchronously spiking neurons of neuronal ensemble corresponding to the astrocyte required for the emergence of astrocyte-induced enhancement of synaptic transmission, $F_{astro}$. Here we do not account for various spatiotemporal properties of gliotransmitter release and astrocyte Ca$^{2+}$ signals evoked by different levels of neuronal activity \citep{Araque2014}. We assume that simultaneous activation of synapses induce multiple Ca$^{2+}$ events at different processes of astrocyte, which are spatially and temporally integrated and result in the generation of a global, long lasting Ca$^{2+}$ elevation \citep{Bindocci2017} that can affect synaptic transmission in the territory of individual astrocytes. For this purpose, parameters $F_{act}$ and $F_{astro}$ estimate the correlation level of synapses activity in the model. The optimal range of $F_{astro}$ for correlation of recalled pattern is [40-60]\% (Fig. \ref{fig_accuracy}b). Smaller values of the parameter, $F_{astro}$, lead to the effect of astrocyte-induced synchronization initiated even by non-stimulus specific uncorrelated noise activity in a small ensemble of neurons. On the contrary, the use of larger $F_{astro}$ values implies that a highly correlated activity of almost all neurons located in the territory of a given astrocyte is required for existence of astrocytic modulation of synapses. Hence, the neuron-astrocytic network can not perform a correct recall of noisy cue. Another point is that the Figure \ref{fig_accuracy}b was obtained for training set with a low 5\% noise level and did not reveal the dependence of the correlation of recalled pattern on the parameter $F_{act}$. We studied the influence of the parameter $F_{act}$ on the correlation in the simulations with different noise in training samples (Fig. \ref{fig_accuracy}c). For lower noise level in training samples, the network memorises items regardless of the value of the parameter, $F_{act}$. Increasing the level of noise in training samples for small values of the parameter, $F_{act}$, leads to the Ca$^{2+}$ elevation in randomly distributed astrocytes first, and then in the whole astrocytic layer. Nevertheless, such nonstimulus-specific astrocytic activation can result 
in high correlation of recalled pattern because of the moderate noise level in the cue and optimally chosen value of the parameter $F_{astro}$. On the contrary, for larger value of the $F_{act} >85$\% Ca$^{2+}$ signal in astrocyte can be evoked only by relatively "clean" sample with a small percentage of noise $<5$\%, therefore increasing the level of noise in training samples results in poor correlation (Fig. \ref{fig_accuracy}c). We found that range of $F_{act} = 80-85 \%$ was optimal for performing the WM tasks by neuron-astrocyte network. In this range astrocyte activations were stimulus-specific and the astrocyte layer could memorise training samples with a low noise level. Note that recent experiments in the investigation of Ca$^{2+}$ activity in astrocytes with precise spatial-temporal resolutions \citep{Bindocci2017} revealed that global calcium event originated from the multiple foci on the most structural part of astrocyte periphery depends on the synchronous neuronal discharges.

Next, we studied the influence of synaptic connectivity architecture in the neural network, specifically the number, weight, and distribution of synaptic connections, on the correlation in the multi-item WM task (Fig.\ref{fig_accuracy_2}). The minimal number of synaptic connections, $N_\text{out}$, required for existence of cued recall is 20 (Fig.\ref{fig_accuracy_2}a). A smaller number of connections is not enough to activate all the neurons from stimulus-specific population. Simultaneous increase of weights and number of connections induces the generation of large synaptic currents resulting in self-sustained overactivation of neuron-astrocyte network. Therefore there exists the optimal range of synaptic weight values to ensure high correlation. We found that for our model this range is $\eta \in [0.005-0.05]$.  Figure \ref{fig_accuracy_2}b illustrates the dependence of correlation on the number of output synaptic connections and their distribution. The smaller the parameter $\lambda$ from Eq.(\ref{eq:Radius}), the lower the probability of long-distance connections. The highest correlation was observed for local connections, $\lambda < 7$, due to the fact that short-range connections do not lead to blurring of the pattern retrieval boundaries in the neural network. Figure \ref{fig_accuracy_2}b also determined the optimal range for the number of synaptic connections: $N_\text{out} \in [25, 55]$.    

The key parameters, which determine the WM capacity in the proposed neuron-astrocyte network, are the duration of calcium signals in astrocyte and duration of the astrocyte-induced modulation of synaptic transmission. Duration of astrocytic calcium elevations is determined by the intrinsic mechanisms of the IP$_3$-evoked  Ca$^{2+}$-induced Ca$^{2+}$ release from the astrocyte endoplasmic reticulum stores, which is described by the biophysical model \citep{Li1994} used in this study. Fragmentary experimental data on duration of the gliotransmitter-induced modulation of synaptic transmission shows that the short-lived version of this modulation lasts from fractions of a second to few minutes, while long-term plasticity can last for tens of minutes (for review see \citep{DePitt2016}). 

\begin{figure}
\centering
\includegraphics[width=0.9\columnwidth]{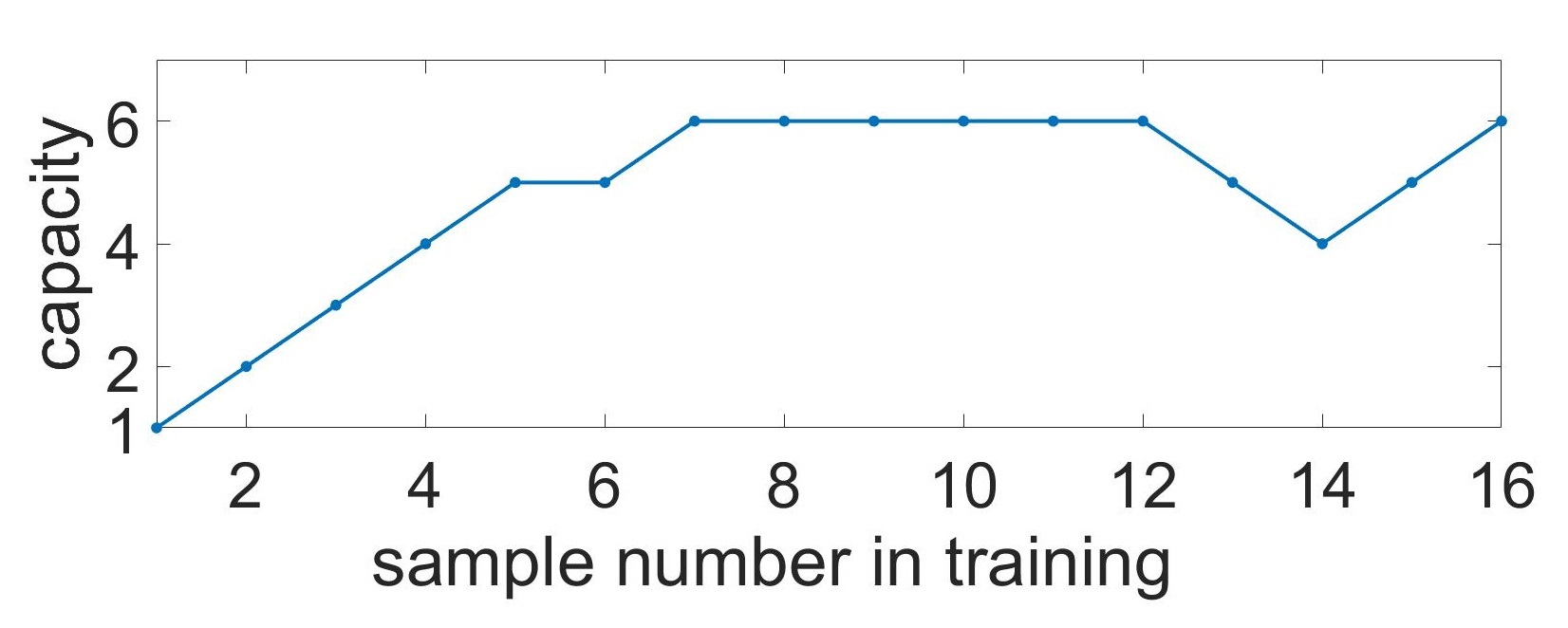}
\caption{Capacity of the multi-item WM in neuron-astrocyte network. Capacity as a function of the sample number in training. The number of images with correlation of recalled pattern higher than 90\% is shown.}\label{fig_capacity}
\end{figure}

To characterise memory capacity, we subjected it to longer trains of samples. Samples were applied to neuronal ensembles with average $35.2$\% overlapping in population. To check the memorisation, we presented cue items in the reverse order compared to learning mode (e.g. learning: $0,1,2,...,7,8$; test: $8,7,...,2,1,0$). The number of items with a correlation of recalled pattern higher than 90\% indicated the capacity of the system. Figure \ref{fig_capacity} shows the capacity as a function of the sample number in the training sequence. For a chosen set of parameters, the capacity of WM ranges from five to six. Note, that such a limited capacity coincides with psychological studies indicating that human ability to keep information in readily accessible WM is limited, ranging between three and five items for most healthy people \citep{Cowan2010}. We investigated the influence of the duration of astrocyte-induced modulation of synaptic transmission, $\tau_\text{astro}$, on the capacity. The number of items stored in the system memory is maximum for parameter values  in the range: $\tau_\text{astro} \in [60, 360]$ ms. For small values of $\tau_\text{astro}$ excitation does not have time to spread to the stimulus-specific neural population; for long astrocytic modulation the different items in cued recalls interfered with each other.
 
\begin{figure}
\centering
\includegraphics[width=0.9\columnwidth]{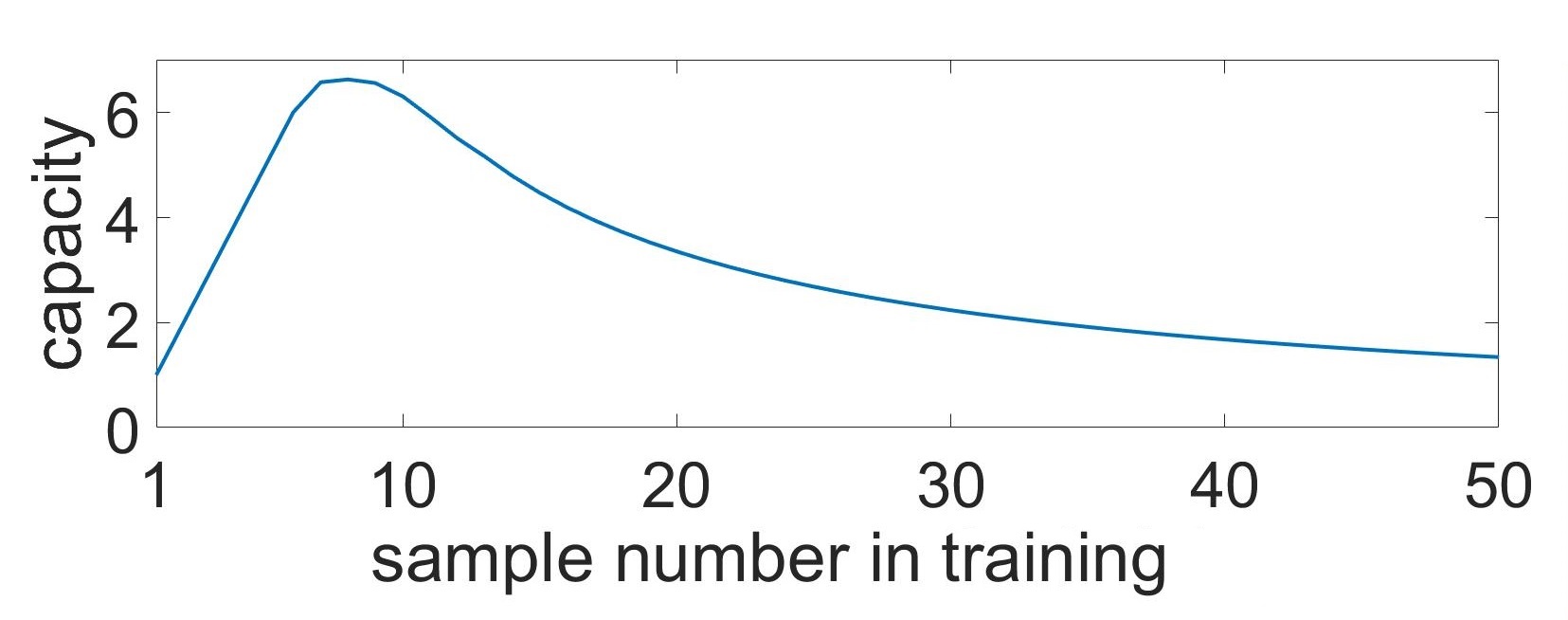}
\caption{Capacity of the multi-item WM in neuron-astrocyte network in the case of non-overlapping neuronal populations obtained analytically. Capacity as a function of the sample number in training sequence.}\label{fig_capacity_analytical}
\end{figure}

In the case of non-overlapping neuronal ensembles, the WM capacity is unequivocally determined by the duration of the astrocytic calcium signal and can be obtained analytically. 
 We estimated the Ca$^{2+}$ signal duration in $\tau_{Ca} = 3.8$ sec. We consider the case of training in $K$ samples of a fixed order: $1,2,...,K$. During the test, we look at a permutation $p$ consisting of $K$ patterns. For the permutation, we estimate the number of correctly recalled patterns, $K_p$. Pattern is considered correctly recalled if no more than $\tau_\text{astro}$ has passed since its presentation. The average capacity, $C$, for this case is defined as the average number of correctly recalled patterns over all possible permutations $p$. According to this description the average capacity can be calculated by the following equation:

\begin{equation}
\label{eq:capacity}
\begin{aligned}
&t_{train}^{i}=i \cdot \tau_{11} + (i - 1) \tau_{12},\\
&t_{test}^{j,i}=t_{train}^{i} + \tau_{shift} + j \cdot \tau_{21} + (j - 1) \tau_{22},\\
&C=\frac{1}{K!}\sum_{p}\sum_{1}^{K}\left[t_{test}^{p_i, i} - t_{train}^{i} < \tau_{Ca}\right]=\\
&\frac{1}{K!}\sum_{1}^{K}\sum_{p}\left[t_{test}^{p_i, i} - t_{train}^{i} < \tau_{Ca}\right]=\\
&\frac{(K-1)!}{K!}\sum_{1}^{K}\sum_{j=1}^{K}\left[t_{test}^{j, i} - t_{train}^{i} < \tau_{Ca}\right],\\
\end{aligned}
\end{equation}
where $\tau_{11}$ - train sample duration, $\tau_{12}$ - duration between train samples, $\tau_{21}$ - test sample duration, $\tau_{22}$ - duration between test samples, $\tau_{shift}$ - delay between train and test, $\tau_{Ca}$ - calcium event duration, $t_{train}^{i}$ - time of i-th train sample finishes, $t_{test}^{j,i}$. Figure \ref{fig_capacity_analytical} shows the capacity as a function of the sample number, $K$. After it reaches a maximum of 6.6 for 8 samples, the capacity begins to decrease monotonically, while the number of samples increases.

\section{Discussion}

We proposed a biologycally motivated spiking neuron network model accompanied by astrocytes that demonstrates the working memory formation. The model acts at multiple timescales: at a millisecond scale of firing neurons and the second scale of calcium dynamics in astrocyte. Neuronal network consists of randomly sparsely connected excitatory spiking neurons with non-plastic synapses. Astrocyte induced activity-dependent short-term synaptic plasticity results in local spatial synchronization in neuronal ensembles. WM realised by such astrocytic modulation is characterised by one-shot learning and is maintained during seconds. The astrocyte influence on the synaptic connections during the elevation of calcium concentration implements Hebbian-like synaptic plasticity differentiating between specific and non-specific activations. Note that the proposed model is crucially different from the attractor-based network memory models \citep{Hopfield1982,Amit1995,Wang2001,Wimmer2014} and works similarly to WM models based on synaptic plasticity \citep{Mongillo2008,Manohar2019,Lundqvist2011,Mi2017}. In particular, in its functionality, the model is quite close to short-term associative (Hebbian) synaptic facilitation \citep{Fiebig2016,Sandberg2003}. 

\begin{figure}
\centering
\includegraphics[width=1.0\columnwidth]{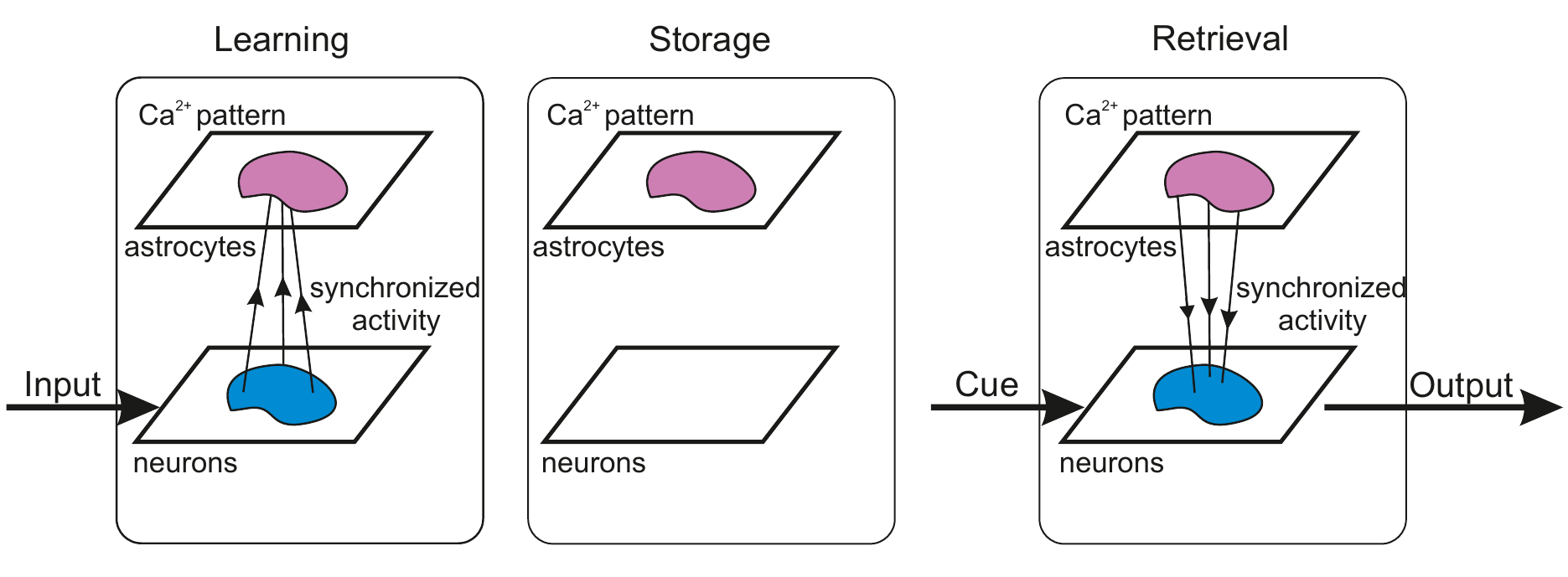}
\caption{Concept of WM operation in spiking neuron network model accompanied by astrocytes.}\label{fig_concept}
\end{figure}

The concept of our WM model operation is schematically summarised in Fig. \ref{fig_concept}. Composed of two building blocks, e.g. fast-spiking neurons and slow astrocytes, the proposed memory architecture eventually demonstrated synergetic functionality in loading information and its readout by the neuronal block and storage implemented by the astrocytes. In contrast with solely neuronal circuit models where memory is encoded in synaptic connections and their plasticity, which inevitably leads to the problem of overlapping, our model splits functionality using astrocytes as a pool for stored patterns. Even with significant overlaps, they can be successfully retrieved due to coherent synaptic modulations by the astrocytes and synchronous neuron firing, which provide the selectivity. When the memory is maintained at the time scale of calcium elevation in astrocytes, the synapses are not specifically modulated and theoretically can be employed for other tasks. Note also that the memory is transient with no long-lasting changes in structure and parameters. Thus, new information can be successfully uploaded and stored without interference with traces of previously memorised information. Complete memory overwrite interval is estimated at several seconds and is also defined by the duration of astrocyte activations.            

The proposed model clearly confirms the theoretical hypothesis that astrocytic modulation of synaptic transmission can be involved in the formation of functional cortical WM. We show that the NMDAR-mediated potentiation of excitatory synapses induced by the D-serine released from astrocyte in PFC could serve as a possible molecular mechanism for WM. Stages of the multi-item WM, which include loading, storage, and cued recall, manifest in brief oscillatory bursts, which are functionally similar to WM activity in nonhuman primate PFC \citep{Lundqvist2016} rather than sustained neuron spiking. Interestingly, the activity of the neuron-astrocyte network corresponding to memorised pattern exhibits a sufficient degree of stability, which ensures memory retention despite the presence of significant overlaps in the stimulus-specific subnetworks. 

Needless, to say that astrocyte-induced modulation of synaptic transmission proposed in this study as a mechanism for WM organisation does not exclude but rather complements other synaptic and neural plasticity mechanisms (fast Hebbian synaptic plasticity/short-term synaptic plasticity, facilitation, augmentation, dendritic voltage bistability, etc.) and may well act in parallel to them.

On the one hand, there has been much experimental evidence that astrocytes contribute to synaptic plasticity, coordination of neural network oscillatory activity, and memory function  \citep{Santello2019}. It was shown recently that astrocytic impact is circuit-specific \citep{Martin2015} and stimulus-specific \citep{Mariotti2018}. Improved Ca$^{2+}$ imaging approaches have identified a spatiotemporal diversity of astrocytic signals that may underlie the capacity of astrocytes to encode and process different patterns of activation \citep{Bindocci2017,Stobart2018}. Besides, the temporal scale of the astrocytic calcium dynamics and dynamics of the neuron-astrocyte bidirectional communication including the effects of astrocytic influence on synaptic plasticity fit very well in timing required in WM processes.

On the other hand, the ongoing intense debate about principles of WM organisation challenges the canonical theory of persistent delay activity in network attractors with recurrent excitation \citep{Bouchacourt2019} and offer alternative models incorporating different biophysical network mechanisms of WM  \citep{Lundqvist2018,Barak2014}. The principal reasons for such a debate is the reexamination of experimental data, which shows a large heterogeneity in the delay neuronal activity during WM tasks \citep{Stokes2013}. A non-classical WM model includes the short-term synaptic plasticity \citep{Mongillo2008, Hansel2013}, the balance of inhibition and excitation \citep{Boerlin2013}, the NMDA currents affecting on the neuronal excitability \citep{Durstewitz2009}, and other parameters. These models, however, have a number of shortcomings: inability to describe encoding of novel associations in synaptic facilitation-based model; unclear mechanisms for achieving precise tuning of recurrent excitation and inhibition; the time constant of the NMDA receptor is appropriate to maintain memories for 1–5 s, but not longer.The investigation of the synaptic mechanisms underlying WM is an ongoing process. Therefore, incorporation of astrocytes as spatiotemporal integrator and modulator of synaptic transmission in neural network models may help advance the theoretical framework of WM encoding and maintenance mechanisms.

Future research direction in the framework of the proposed WM model in the neuron-astrocyte network will be focused on the interplay of excitation and inhibition that can stabilise WM [100]; the effects of synaptic plasticity \citep{Boerlin2013}; the effects of synaptic plasticity \citep{Hansel2013,Mongillo2008} namely associative short-term potentiation (a fast-expressing form of Hebbian synaptic plasticity) that can provide an encoding of novel associations \citep{Fiebig2016}; and on the structure of the cortical microcircuit reflecting columnar organisation of the neocortex.

	This research was supported by the Ministry of Science and Higher Education of the Russian Federation (project No. 075-15-2020-808, No. 0729-2020-0061). SG thanks the RFBR (grant No. 20-32-70081).

\bibliography{refs}

\begin{thebibliography}{107}
\expandafter\ifx\csname natexlab\endcsname\relax\def\natexlab#1{#1}\fi
\providecommand{\url}[1]{\texttt{#1}}
\providecommand{\href}[2]{#2}
\providecommand{\path}[1]{#1}
\providecommand{\DOIprefix}{doi:}
\providecommand{\ArXivprefix}{arXiv:}
\providecommand{\URLprefix}{URL: }
\providecommand{\Pubmedprefix}{pmid:}
\providecommand{\doi}[1]{\href{http://dx.doi.org/#1}{\path{#1}}}
\providecommand{\Pubmed}[1]{\href{pmid:#1}{\path{#1}}}
\providecommand{\bibinfo}[2]{#2}
\ifx\xfnm\relax \def\xfnm[#1]{\unskip,\space#1}\fi
%Type = Article
\bibitem[{Allen and Eroglu(2017)}]{Allen2017}
\bibinfo{author}{Allen, N.J.}, \bibinfo{author}{Eroglu, C.},
  \bibinfo{year}{2017}.
\newblock \bibinfo{title}{Cell biology of astrocyte-synapse interactions}.
\newblock \bibinfo{journal}{Neuron} \bibinfo{volume}{96},
  \bibinfo{pages}{697--708}.
\newblock \URLprefix \url{https://doi.org/10.1016/j.neuron.2017.09.056},
  \DOIprefix\doi{10.1016/j.neuron.2017.09.056}.
%Type = Article
\bibitem[{Amit(2003)}]{Amit2003}
\bibinfo{author}{Amit, D.}, \bibinfo{year}{2003}.
\newblock \bibinfo{title}{Multiple-object working memory--a model for
  behavioral performance}.
\newblock \bibinfo{journal}{Cerebral Cortex} \bibinfo{volume}{13},
  \bibinfo{pages}{435--443}.
\newblock \URLprefix \url{https://doi.org/10.1093/cercor/13.5.435},
  \DOIprefix\doi{10.1093/cercor/13.5.435}.
%Type = Article
\bibitem[{Amit et~al.(2013)Amit, Yakovlev and Hochstein}]{Amit2013}
\bibinfo{author}{Amit, Y.}, \bibinfo{author}{Yakovlev, V.},
  \bibinfo{author}{Hochstein, S.}, \bibinfo{year}{2013}.
\newblock \bibinfo{title}{Modeling behavior in different delay match to sample
  tasks in one simple network}.
\newblock \bibinfo{journal}{Frontiers in Human Neuroscience}
  \bibinfo{volume}{7}.
\newblock \URLprefix \url{https://doi.org/10.3389/fnhum.2013.00408},
  \DOIprefix\doi{10.3389/fnhum.2013.00408}.
%Type = Article
\bibitem[{Araque et~al.(2014)Araque, Carmignoto, Haydon, Oliet, Robitaille and
  Volterra}]{Araque2014}
\bibinfo{author}{Araque, A.}, \bibinfo{author}{Carmignoto, G.},
  \bibinfo{author}{Haydon, P.G.}, \bibinfo{author}{Oliet, S.H.},
  \bibinfo{author}{Robitaille, R.}, \bibinfo{author}{Volterra, A.},
  \bibinfo{year}{2014}.
\newblock \bibinfo{title}{Gliotransmitters travel in time and space}.
\newblock \bibinfo{journal}{Neuron} \bibinfo{volume}{81},
  \bibinfo{pages}{728--739}.
\newblock \URLprefix \url{https://doi.org/10.1016/j.neuron.2014.02.007},
  \DOIprefix\doi{10.1016/j.neuron.2014.02.007}.
%Type = Article
\bibitem[{Attwell and Laughlin(2001)}]{Attwell2001}
\bibinfo{author}{Attwell, D.}, \bibinfo{author}{Laughlin, S.B.},
  \bibinfo{year}{2001}.
\newblock \bibinfo{title}{An energy budget for signaling in the grey matter of
  the brain}.
\newblock \bibinfo{journal}{Journal of Cerebral Blood Flow {\&} Metabolism}
  \bibinfo{volume}{21}, \bibinfo{pages}{1133--1145}.
\newblock \URLprefix \url{https://doi.org/10.1097/00004647-200110000-00001},
  \DOIprefix\doi{10.1097/00004647-200110000-00001}.
%Type = Article
\bibitem[{Baddeley(2012)}]{Baddeley2012}
\bibinfo{author}{Baddeley, A.}, \bibinfo{year}{2012}.
\newblock \bibinfo{title}{Working memory: Theories, models, and controversies}.
\newblock \bibinfo{journal}{Annual Review of Psychology} \bibinfo{volume}{63},
  \bibinfo{pages}{1--29}.
\newblock \URLprefix \url{https://doi.org/10.1146/annurev-psych-120710-100422},
  \DOIprefix\doi{10.1146/annurev-psych-120710-100422}.
%Type = Book
\bibitem[{Baddeley(1986)}]{Baddeley1986}
\bibinfo{author}{Baddeley, A.D.}, \bibinfo{year}{1986}.
\newblock \bibinfo{title}{Working memory}.
\newblock \bibinfo{publisher}{Clarendon Press ; New York : Oxford University
  Press}.
%Type = Article
\bibitem[{Barak and Tsodyks(2014)}]{Barak2014}
\bibinfo{author}{Barak, O.}, \bibinfo{author}{Tsodyks, M.},
  \bibinfo{year}{2014}.
\newblock \bibinfo{title}{Working models of working memory}.
\newblock \bibinfo{journal}{Current Opinion in Neurobiology}
  \bibinfo{volume}{25}, \bibinfo{pages}{20--24}.
\newblock \URLprefix \url{https://doi.org/10.1016/j.conb.2013.10.008},
  \DOIprefix\doi{10.1016/j.conb.2013.10.008}.
%Type = Article
\bibitem[{Barak et~al.(2010)Barak, Tsodyks and Romo}]{Barak2010}
\bibinfo{author}{Barak, O.}, \bibinfo{author}{Tsodyks, M.},
  \bibinfo{author}{Romo, R.}, \bibinfo{year}{2010}.
\newblock \bibinfo{title}{Neuronal population coding of parametric working
  memory}.
\newblock \bibinfo{journal}{Journal of Neuroscience} \bibinfo{volume}{30},
  \bibinfo{pages}{9424--9430}.
\newblock \URLprefix \url{https://doi.org/10.1523/jneurosci.1875-10.2010},
  \DOIprefix\doi{10.1523/jneurosci.1875-10.2010}.
%Type = Article
\bibitem[{Bekar et~al.(2008)Bekar, He and Nedergaard}]{Bekar2008}
\bibinfo{author}{Bekar, L.K.}, \bibinfo{author}{He, W.},
  \bibinfo{author}{Nedergaard, M.}, \bibinfo{year}{2008}.
\newblock \bibinfo{title}{Locus coeruleus
  $\alpha$-adrenergic{\textendash}mediated activation of cortical astrocytes in
  vivo}.
\newblock \bibinfo{journal}{Cerebral Cortex} \bibinfo{volume}{18},
  \bibinfo{pages}{2789--2795}.
\newblock \URLprefix \url{https://doi.org/10.1093/cercor/bhn040},
  \DOIprefix\doi{10.1093/cercor/bhn040}.
%Type = Article
\bibitem[{Bergersen et~al.(2011)Bergersen, Morland, Ormel, Rinholm, Larsson,
  Wold, R{\o}e, Stranna, Santello, Bouvier, Ottersen, Volterra and
  Gundersen}]{Bergersen2011}
\bibinfo{author}{Bergersen, L.}, \bibinfo{author}{Morland, C.},
  \bibinfo{author}{Ormel, L.}, \bibinfo{author}{Rinholm, J.},
  \bibinfo{author}{Larsson, M.}, \bibinfo{author}{Wold, J.},
  \bibinfo{author}{R{\o}e, {\AA}.}, \bibinfo{author}{Stranna, A.},
  \bibinfo{author}{Santello, M.}, \bibinfo{author}{Bouvier, D.},
  \bibinfo{author}{Ottersen, O.}, \bibinfo{author}{Volterra, A.},
  \bibinfo{author}{Gundersen, V.}, \bibinfo{year}{2011}.
\newblock \bibinfo{title}{Immunogold detection of l-glutamate and d-serine in
  small synaptic-like microvesicles in adult hippocampal astrocytes}.
\newblock \bibinfo{journal}{Cerebral Cortex} \bibinfo{volume}{22},
  \bibinfo{pages}{1690--1697}.
\newblock \URLprefix \url{https://doi.org/10.1093/cercor/bhr254},
  \DOIprefix\doi{10.1093/cercor/bhr254}.
%Type = Article
\bibitem[{Bindocci et~al.(2017)Bindocci, Savtchouk, Liaudet, Becker, Carriero
  and Volterra}]{Bindocci2017}
\bibinfo{author}{Bindocci, E.}, \bibinfo{author}{Savtchouk, I.},
  \bibinfo{author}{Liaudet, N.}, \bibinfo{author}{Becker, D.},
  \bibinfo{author}{Carriero, G.}, \bibinfo{author}{Volterra, A.},
  \bibinfo{year}{2017}.
\newblock \bibinfo{title}{Three-dimensional ca2$+$imaging advances
  understanding of astrocyte biology}.
\newblock \bibinfo{journal}{Science} \bibinfo{volume}{356},
  \bibinfo{pages}{eaai8185}.
\newblock \URLprefix \url{https://doi.org/10.1126/science.aai8185},
  \DOIprefix\doi{10.1126/science.aai8185}.
%Type = Article
\bibitem[{Boerlin et~al.(2013)Boerlin, Machens and Den{\`{e}}ve}]{Boerlin2013}
\bibinfo{author}{Boerlin, M.}, \bibinfo{author}{Machens, C.K.},
  \bibinfo{author}{Den{\`{e}}ve, S.}, \bibinfo{year}{2013}.
\newblock \bibinfo{title}{Predictive coding of dynamical variables in balanced
  spiking networks}.
\newblock \bibinfo{journal}{{PLoS} Computational Biology} \bibinfo{volume}{9},
  \bibinfo{pages}{e1003258}.
\newblock \URLprefix \url{https://doi.org/10.1371/journal.pcbi.1003258},
  \DOIprefix\doi{10.1371/journal.pcbi.1003258}.
%Type = Article
\bibitem[{Borisyuk et~al.(2013)Borisyuk, Chik, Kazanovich and
  da~Silva~Gomes}]{Borisyuk2013}
\bibinfo{author}{Borisyuk, R.}, \bibinfo{author}{Chik, D.},
  \bibinfo{author}{Kazanovich, Y.}, \bibinfo{author}{da~Silva~Gomes, J.},
  \bibinfo{year}{2013}.
\newblock \bibinfo{title}{Spiking neural network model for memorizing sequences
  with forward and backward recall}.
\newblock \bibinfo{journal}{Biosystems} \bibinfo{volume}{112},
  \bibinfo{pages}{214--223}.
\newblock \URLprefix \url{https://doi.org/10.1016/j.biosystems.2013.03.018},
  \DOIprefix\doi{10.1016/j.biosystems.2013.03.018}.
%Type = Article
\bibitem[{Bouchacourt and Buschman(2019)}]{Bouchacourt2019}
\bibinfo{author}{Bouchacourt, F.}, \bibinfo{author}{Buschman, T.J.},
  \bibinfo{year}{2019}.
\newblock \bibinfo{title}{A flexible model of working memory}.
\newblock \bibinfo{journal}{Neuron} \bibinfo{volume}{103},
  \bibinfo{pages}{147--160.e8}.
\newblock \URLprefix \url{https://doi.org/10.1016/j.neuron.2019.04.020},
  \DOIprefix\doi{10.1016/j.neuron.2019.04.020}.
%Type = Article
\bibitem[{Brunel and Wang(2001)}]{Brunel2001}
\bibinfo{author}{Brunel, N.}, \bibinfo{author}{Wang, X.J.},
  \bibinfo{year}{2001}.
\newblock \bibinfo{title}{Effects of neuromodulation in a cortical network
  model of object working memory dominated by recurrent inhibition}.
\newblock \bibinfo{journal}{Journal of Computational Neuroscience}
  \bibinfo{volume}{11}, \bibinfo{pages}{63--85}.
\newblock \URLprefix \url{https://doi.org/10.1023/a:1011204814320},
  \DOIprefix\doi{10.1023/a:1011204814320}.
%Type = Article
\bibitem[{Chaudhuri and Fiete(2016)}]{Chaudhuri2016}
\bibinfo{author}{Chaudhuri, R.}, \bibinfo{author}{Fiete, I.},
  \bibinfo{year}{2016}.
\newblock \bibinfo{title}{Computational principles of memory}.
\newblock \bibinfo{journal}{Nature Neuroscience} \bibinfo{volume}{19},
  \bibinfo{pages}{394--403}.
\newblock \URLprefix \url{https://doi.org/10.1038/nn.4237},
  \DOIprefix\doi{10.1038/nn.4237}.
%Type = Article
\bibitem[{Chen et~al.(2012)Chen, Sugihara, Sharma, Perea, Petravicz, Le and
  Sur}]{Chen2012}
\bibinfo{author}{Chen, N.}, \bibinfo{author}{Sugihara, H.},
  \bibinfo{author}{Sharma, J.}, \bibinfo{author}{Perea, G.},
  \bibinfo{author}{Petravicz, J.}, \bibinfo{author}{Le, C.},
  \bibinfo{author}{Sur, M.}, \bibinfo{year}{2012}.
\newblock \bibinfo{title}{Nucleus basalis-enabled stimulus-specific plasticity
  in the visual cortex is mediated by astrocytes}.
\newblock \bibinfo{journal}{Proceedings of the National Academy of Sciences}
  \bibinfo{volume}{109}, \bibinfo{pages}{E2832--E2841}.
\newblock \URLprefix \url{https://doi.org/10.1073/pnas.1206557109},
  \DOIprefix\doi{10.1073/pnas.1206557109}.
%Type = Article
\bibitem[{Constantinidis et~al.(2018)Constantinidis, Funahashi, Lee, Murray,
  Qi, Wang and Arnsten}]{Constantinidis2018}
\bibinfo{author}{Constantinidis, C.}, \bibinfo{author}{Funahashi, S.},
  \bibinfo{author}{Lee, D.}, \bibinfo{author}{Murray, J.D.},
  \bibinfo{author}{Qi, X.L.}, \bibinfo{author}{Wang, M.},
  \bibinfo{author}{Arnsten, A.F.}, \bibinfo{year}{2018}.
\newblock \bibinfo{title}{Persistent spiking activity underlies working
  memory}.
\newblock \bibinfo{journal}{The Journal of Neuroscience} \bibinfo{volume}{38},
  \bibinfo{pages}{7020--7028}.
\newblock \URLprefix \url{https://doi.org/10.1523/jneurosci.2486-17.2018},
  \DOIprefix\doi{10.1523/jneurosci.2486-17.2018}.
%Type = Article
\bibitem[{Conway et~al.(2003)Conway, Kane and Engle}]{Conway2003}
\bibinfo{author}{Conway, A.R.}, \bibinfo{author}{Kane, M.J.},
  \bibinfo{author}{Engle, R.W.}, \bibinfo{year}{2003}.
\newblock \bibinfo{title}{Working memory capacity and its relation to general
  intelligence}.
\newblock \bibinfo{journal}{Trends in Cognitive Sciences} \bibinfo{volume}{7},
  \bibinfo{pages}{547--552}.
\newblock \URLprefix \url{https://doi.org/10.1016/j.tics.2003.10.005},
  \DOIprefix\doi{10.1016/j.tics.2003.10.005}.
%Type = Article
\bibitem[{Cowan(2010)}]{Cowan2010}
\bibinfo{author}{Cowan, N.}, \bibinfo{year}{2010}.
\newblock \bibinfo{title}{The magical mystery four}.
\newblock \bibinfo{journal}{Current Directions in Psychological Science}
  \bibinfo{volume}{19}, \bibinfo{pages}{51--57}.
\newblock \URLprefix \url{https://doi.org/10.1177/0963721409359277},
  \DOIprefix\doi{10.1177/0963721409359277}.
%Type = Article
\bibitem[{D'Esposito and Postle(2015)}]{DEsposito2015}
\bibinfo{author}{D'Esposito, M.}, \bibinfo{author}{Postle, B.R.},
  \bibinfo{year}{2015}.
\newblock \bibinfo{title}{The cognitive neuroscience of working memory}.
\newblock \bibinfo{journal}{Annual Review of Psychology} \bibinfo{volume}{66},
  \bibinfo{pages}{115--142}.
\newblock \URLprefix \url{https://doi.org/10.1146/annurev-psych-010814-015031},
  \DOIprefix\doi{10.1146/annurev-psych-010814-015031}.
%Type = Book
\bibitem[{DJ(1995)}]{Amit1995}
\bibinfo{author}{DJ, A.}, \bibinfo{year}{1995}.
\newblock \bibinfo{title}{Modeling brain functions: the world of attractor
  neural networks}.
\newblock \bibinfo{publisher}{Cambridge, UK: Cambridge U}.
%Type = Article
\bibitem[{Durkee and Araque(2019)}]{Durkee2019}
\bibinfo{author}{Durkee, C.A.}, \bibinfo{author}{Araque, A.},
  \bibinfo{year}{2019}.
\newblock \bibinfo{title}{Diversity and specificity of
  astrocyte{\textendash}neuron communication}.
\newblock \bibinfo{journal}{Neuroscience} \bibinfo{volume}{396},
  \bibinfo{pages}{73--78}.
\newblock \URLprefix \url{https://doi.org/10.1016/j.neuroscience.2018.11.010},
  \DOIprefix\doi{10.1016/j.neuroscience.2018.11.010}.
%Type = Article
\bibitem[{Durstewitz(2009)}]{Durstewitz2009}
\bibinfo{author}{Durstewitz, D.}, \bibinfo{year}{2009}.
\newblock \bibinfo{title}{Implications of synaptic biophysics for recurrent
  network dynamics and active memory}.
\newblock \bibinfo{journal}{Neural Networks} \bibinfo{volume}{22},
  \bibinfo{pages}{1189--1200}.
\newblock \URLprefix \url{https://doi.org/10.1016/j.neunet.2009.07.016},
  \DOIprefix\doi{10.1016/j.neunet.2009.07.016}.
%Type = Article
\bibitem[{Erickson et~al.(2010)Erickson, Maramara and Lisman}]{Erickson2010}
\bibinfo{author}{Erickson, M.A.}, \bibinfo{author}{Maramara, L.A.},
  \bibinfo{author}{Lisman, J.}, \bibinfo{year}{2010}.
\newblock \bibinfo{title}{A single brief burst induces (glur)1-dependent
  associative short-term potentiation: A potential mechanism for short-term
  memory}.
\newblock \bibinfo{journal}{Journal of Cognitive Neuroscience}
  \bibinfo{volume}{22}, \bibinfo{pages}{2530--2540}.
\newblock \URLprefix \url{https://doi.org/10.1162/jocn.2009.21375},
  \DOIprefix\doi{10.1162/jocn.2009.21375}.
%Type = Article
\bibitem[{Esir et~al.(2018)Esir, Gordleeva, Simonov, Pisarchik and
  Kazantsev}]{Esir2018}
\bibinfo{author}{Esir, P.M.}, \bibinfo{author}{Gordleeva, S.Y.},
  \bibinfo{author}{Simonov, A.Y.}, \bibinfo{author}{Pisarchik, A.N.},
  \bibinfo{author}{Kazantsev, V.B.}, \bibinfo{year}{2018}.
\newblock \bibinfo{title}{Conduction delays can enhance formation of up and
  down states in spiking neuronal networks}.
\newblock \bibinfo{journal}{Physical Review E} \bibinfo{volume}{98}.
\newblock \URLprefix \url{https://doi.org/10.1103/physreve.98.052401},
  \DOIprefix\doi{10.1103/physreve.98.052401}.
%Type = Article
\bibitem[{Fiebig and Lansner(2016)}]{Fiebig2016}
\bibinfo{author}{Fiebig, F.}, \bibinfo{author}{Lansner, A.},
  \bibinfo{year}{2016}.
\newblock \bibinfo{title}{A spiking working memory model based on hebbian
  short-term potentiation}.
\newblock \bibinfo{journal}{The Journal of Neuroscience} \bibinfo{volume}{37},
  \bibinfo{pages}{83--96}.
\newblock \URLprefix \url{https://doi.org/10.1523/jneurosci.1989-16.2016},
  \DOIprefix\doi{10.1523/jneurosci.1989-16.2016}.
%Type = Article
\bibitem[{Fossat et~al.(2011)Fossat, Turpin, Sacchi, Dulong, Shi, Rivet,
  Sweedler, Pollegioni, Millan, Oliet and Mothet}]{Fossat2011}
\bibinfo{author}{Fossat, P.}, \bibinfo{author}{Turpin, F.R.},
  \bibinfo{author}{Sacchi, S.}, \bibinfo{author}{Dulong, J.},
  \bibinfo{author}{Shi, T.}, \bibinfo{author}{Rivet, J.M.},
  \bibinfo{author}{Sweedler, J.V.}, \bibinfo{author}{Pollegioni, L.},
  \bibinfo{author}{Millan, M.J.}, \bibinfo{author}{Oliet, S.H.},
  \bibinfo{author}{Mothet, J.P.}, \bibinfo{year}{2011}.
\newblock \bibinfo{title}{Glial d-serine gates {NMDA} receptors at excitatory
  synapses in prefrontal cortex}.
\newblock \bibinfo{journal}{Cerebral Cortex} \bibinfo{volume}{22},
  \bibinfo{pages}{595--606}.
\newblock \URLprefix \url{https://doi.org/10.1093/cercor/bhr130},
  \DOIprefix\doi{10.1093/cercor/bhr130}.
%Type = Article
\bibitem[{Frans{\'{e}}n et~al.(2006)Frans{\'{e}}n, Tahvildari, Egorov, Hasselmo
  and Alonso}]{Fransn2006}
\bibinfo{author}{Frans{\'{e}}n, E.}, \bibinfo{author}{Tahvildari, B.},
  \bibinfo{author}{Egorov, A.V.}, \bibinfo{author}{Hasselmo, M.E.},
  \bibinfo{author}{Alonso, A.A.}, \bibinfo{year}{2006}.
\newblock \bibinfo{title}{Mechanism of graded persistent cellular activity of
  entorhinal cortex layer v neurons}.
\newblock \bibinfo{journal}{Neuron} \bibinfo{volume}{49},
  \bibinfo{pages}{735--746}.
\newblock \URLprefix \url{https://doi.org/10.1016/j.neuron.2006.01.036},
  \DOIprefix\doi{10.1016/j.neuron.2006.01.036}.
%Type = Article
\bibitem[{Fujisawa et~al.(2008)Fujisawa, Amarasingham, Harrison and
  Buzs{\'{a}}ki}]{Fujisawa2008}
\bibinfo{author}{Fujisawa, S.}, \bibinfo{author}{Amarasingham, A.},
  \bibinfo{author}{Harrison, M.T.}, \bibinfo{author}{Buzs{\'{a}}ki, G.},
  \bibinfo{year}{2008}.
\newblock \bibinfo{title}{Behavior-dependent short-term assembly dynamics in
  the medial prefrontal cortex}.
\newblock \bibinfo{journal}{Nature Neuroscience} \bibinfo{volume}{11},
  \bibinfo{pages}{823--833}.
\newblock \URLprefix \url{https://doi.org/10.1038/nn.2134},
  \DOIprefix\doi{10.1038/nn.2134}.
%Type = Article
\bibitem[{Funahashi(2017)}]{Funahashi2017}
\bibinfo{author}{Funahashi, S.}, \bibinfo{year}{2017}.
\newblock \bibinfo{title}{Working memory in the prefrontal cortex}.
\newblock \bibinfo{journal}{Brain Sciences} \bibinfo{volume}{7},
  \bibinfo{pages}{49}.
\newblock \URLprefix \url{https://doi.org/10.3390/brainsci7050049},
  \DOIprefix\doi{10.3390/brainsci7050049}.
%Type = Article
\bibitem[{Funahashi et~al.(1989)Funahashi, Bruce and
  Goldman-Rakic}]{Funahashi1989}
\bibinfo{author}{Funahashi, S.}, \bibinfo{author}{Bruce, C.J.},
  \bibinfo{author}{Goldman-Rakic, P.S.}, \bibinfo{year}{1989}.
\newblock \bibinfo{title}{Mnemonic coding of visual space in the monkey's
  dorsolateral prefrontal cortex}.
\newblock \bibinfo{journal}{Journal of Neurophysiology} \bibinfo{volume}{61},
  \bibinfo{pages}{331--349}.
\newblock \URLprefix \url{https://doi.org/10.1152/jn.1989.61.2.331},
  \DOIprefix\doi{10.1152/jn.1989.61.2.331}.
%Type = Article
\bibitem[{Fuster and Alexander(1971)}]{Fuster1971}
\bibinfo{author}{Fuster, J.M.}, \bibinfo{author}{Alexander, G.E.},
  \bibinfo{year}{1971}.
\newblock \bibinfo{title}{Neuron activity related to short-term memory}.
\newblock \bibinfo{journal}{Science} \bibinfo{volume}{173},
  \bibinfo{pages}{652--654}.
\newblock \URLprefix \url{https://doi.org/10.1126/science.173.3997.652},
  \DOIprefix\doi{10.1126/science.173.3997.652}.
%Type = Article
\bibitem[{Ganguli and Latham(2009)}]{Ganguli2009}
\bibinfo{author}{Ganguli, S.}, \bibinfo{author}{Latham, P.},
  \bibinfo{year}{2009}.
\newblock \bibinfo{title}{Feedforward to the past: The relation between
  neuronal connectivity, amplification, and short-term memory}.
\newblock \bibinfo{journal}{Neuron} \bibinfo{volume}{61},
  \bibinfo{pages}{499--501}.
\newblock \URLprefix \url{https://doi.org/10.1016/j.neuron.2009.02.006},
  \DOIprefix\doi{10.1016/j.neuron.2009.02.006}.
%Type = Article
\bibitem[{Goldman(2009)}]{Goldman2009}
\bibinfo{author}{Goldman, M.S.}, \bibinfo{year}{2009}.
\newblock \bibinfo{title}{Memory without feedback in a neural network}.
\newblock \bibinfo{journal}{Neuron} \bibinfo{volume}{61},
  \bibinfo{pages}{621--634}.
\newblock \URLprefix \url{https://doi.org/10.1016/j.neuron.2008.12.012},
  \DOIprefix\doi{10.1016/j.neuron.2008.12.012}.
%Type = Article
\bibitem[{Goldman-Rakic(1995)}]{GoldmanRakic1995}
\bibinfo{author}{Goldman-Rakic, P.}, \bibinfo{year}{1995}.
\newblock \bibinfo{title}{Cellular basis of working memory}.
\newblock \bibinfo{journal}{Neuron} \bibinfo{volume}{14},
  \bibinfo{pages}{477--485}.
\newblock \URLprefix \url{https://doi.org/10.1016/0896-6273(95)90304-6},
  \DOIprefix\doi{10.1016/0896-6273(95)90304-6}.
%Type = Article
\bibitem[{Gordleeva et~al.(2019)Gordleeva, Ermolaeva, Kastalskiy and
  Kazantsev}]{Gordleeva2019}
\bibinfo{author}{Gordleeva, S.Y.}, \bibinfo{author}{Ermolaeva, A.V.},
  \bibinfo{author}{Kastalskiy, I.A.}, \bibinfo{author}{Kazantsev, V.B.},
  \bibinfo{year}{2019}.
\newblock \bibinfo{title}{Astrocyte as spatiotemporal integrating detector of
  neuronal activity}.
\newblock \bibinfo{journal}{Frontiers in Physiology} \bibinfo{volume}{10}.
\newblock \URLprefix \url{https://doi.org/10.3389/fphys.2019.00294},
  \DOIprefix\doi{10.3389/fphys.2019.00294}.
%Type = Article
\bibitem[{Gordleeva et~al.(2018)Gordleeva, Lebedev, Rumyantseva and
  Kazantsev}]{Gordleeva2018}
\bibinfo{author}{Gordleeva, S.Y.}, \bibinfo{author}{Lebedev, S.A.},
  \bibinfo{author}{Rumyantseva, M.A.}, \bibinfo{author}{Kazantsev, V.B.},
  \bibinfo{year}{2018}.
\newblock \bibinfo{title}{Astrocyte as a detector of synchronous events of a
  neural network}.
\newblock \bibinfo{journal}{{JETP} Letters} \bibinfo{volume}{107},
  \bibinfo{pages}{440--445}.
\newblock \URLprefix \url{https://doi.org/10.1134/s0021364018070032},
  \DOIprefix\doi{10.1134/s0021364018070032}.
%Type = Article
\bibitem[{Gordleeva et~al.(2012)Gordleeva, Stasenko, Semyanov, Dityatev and
  Kazantsev}]{Gordleeva2012}
\bibinfo{author}{Gordleeva, S.Y.}, \bibinfo{author}{Stasenko, S.V.},
  \bibinfo{author}{Semyanov, A.V.}, \bibinfo{author}{Dityatev, A.E.},
  \bibinfo{author}{Kazantsev, V.B.}, \bibinfo{year}{2012}.
\newblock \bibinfo{title}{Bi-directional astrocytic regulation of neuronal
  activity within a network}.
\newblock \bibinfo{journal}{Frontiers in Computational Neuroscience}
  \bibinfo{volume}{6}.
\newblock \URLprefix \url{https://doi.org/10.3389/fncom.2012.00092},
  \DOIprefix\doi{10.3389/fncom.2012.00092}.
%Type = Article
\bibitem[{Halassa et~al.(2007)Halassa, Fellin, Takano, Dong and
  Haydon}]{Halassa2007}
\bibinfo{author}{Halassa, M.M.}, \bibinfo{author}{Fellin, T.},
  \bibinfo{author}{Takano, H.}, \bibinfo{author}{Dong, J.H.},
  \bibinfo{author}{Haydon, P.G.}, \bibinfo{year}{2007}.
\newblock \bibinfo{title}{Synaptic islands defined by the territory of a single
  astrocyte}.
\newblock \bibinfo{journal}{Journal of Neuroscience} \bibinfo{volume}{27},
  \bibinfo{pages}{6473--6477}.
\newblock \URLprefix \url{https://doi.org/10.1523/jneurosci.1419-07.2007},
  \DOIprefix\doi{10.1523/jneurosci.1419-07.2007}.
%Type = Article
\bibitem[{Han et~al.(2012)Han, Kesner, Metna-Laurent, Duan, Xu, Georges, Koehl,
  Abrous, Mendizabal-Zubiaga, Grandes, Liu, Bai, Wang, Xiong, Ren, Marsicano
  and Zhang}]{Han2012}
\bibinfo{author}{Han, J.}, \bibinfo{author}{Kesner, P.},
  \bibinfo{author}{Metna-Laurent, M.}, \bibinfo{author}{Duan, T.},
  \bibinfo{author}{Xu, L.}, \bibinfo{author}{Georges, F.},
  \bibinfo{author}{Koehl, M.}, \bibinfo{author}{Abrous, D.N.},
  \bibinfo{author}{Mendizabal-Zubiaga, J.}, \bibinfo{author}{Grandes, P.},
  \bibinfo{author}{Liu, Q.}, \bibinfo{author}{Bai, G.}, \bibinfo{author}{Wang,
  W.}, \bibinfo{author}{Xiong, L.}, \bibinfo{author}{Ren, W.},
  \bibinfo{author}{Marsicano, G.}, \bibinfo{author}{Zhang, X.},
  \bibinfo{year}{2012}.
\newblock \bibinfo{title}{Acute cannabinoids impair working memory through
  astroglial {CB}1 receptor modulation of hippocampal {LTD}}.
\newblock \bibinfo{journal}{Cell} \bibinfo{volume}{148},
  \bibinfo{pages}{1039--1050}.
\newblock \URLprefix \url{https://doi.org/10.1016/j.cell.2012.01.037},
  \DOIprefix\doi{10.1016/j.cell.2012.01.037}.
%Type = Article
\bibitem[{Hansel and Mato(2013)}]{Hansel2013}
\bibinfo{author}{Hansel, D.}, \bibinfo{author}{Mato, G.}, \bibinfo{year}{2013}.
\newblock \bibinfo{title}{Short-term plasticity explains irregular persistent
  activity in working memory tasks}.
\newblock \bibinfo{journal}{Journal of Neuroscience} \bibinfo{volume}{33},
  \bibinfo{pages}{133--149}.
\newblock \URLprefix \url{https://doi.org/10.1523/jneurosci.3455-12.2013},
  \DOIprefix\doi{10.1523/jneurosci.3455-12.2013}.
%Type = Article
\bibitem[{Hempel et~al.(2000)Hempel, Hartman, Wang, Turrigiano and
  Nelson}]{Hempel2000}
\bibinfo{author}{Hempel, C.M.}, \bibinfo{author}{Hartman, K.H.},
  \bibinfo{author}{Wang, X.J.}, \bibinfo{author}{Turrigiano, G.G.},
  \bibinfo{author}{Nelson, S.B.}, \bibinfo{year}{2000}.
\newblock \bibinfo{title}{Multiple forms of short-term plasticity at excitatory
  synapses in rat medial prefrontal cortex}.
\newblock \bibinfo{journal}{Journal of Neurophysiology} \bibinfo{volume}{83},
  \bibinfo{pages}{3031--3041}.
\newblock \URLprefix \url{https://doi.org/10.1152/jn.2000.83.5.3031},
  \DOIprefix\doi{10.1152/jn.2000.83.5.3031}.
%Type = Article
\bibitem[{Henneberger et~al.(2010)Henneberger, Papouin, Oliet and
  Rusakov}]{Henneberger2010}
\bibinfo{author}{Henneberger, C.}, \bibinfo{author}{Papouin, T.},
  \bibinfo{author}{Oliet, S.H.R.}, \bibinfo{author}{Rusakov, D.A.},
  \bibinfo{year}{2010}.
\newblock \bibinfo{title}{Long-term potentiation depends on release of d-serine
  from astrocytes}.
\newblock \bibinfo{journal}{Nature} \bibinfo{volume}{463},
  \bibinfo{pages}{232--236}.
\newblock \URLprefix \url{https://doi.org/10.1038/nature08673},
  \DOIprefix\doi{10.1038/nature08673}.
%Type = Article
\bibitem[{Hopfield(1982)}]{Hopfield1982}
\bibinfo{author}{Hopfield, J.J.}, \bibinfo{year}{1982}.
\newblock \bibinfo{title}{Neural networks and physical systems with emergent
  collective computational abilities.}
\newblock \bibinfo{journal}{Proceedings of the National Academy of Sciences}
  \bibinfo{volume}{79}, \bibinfo{pages}{2554--2558}.
\newblock \URLprefix \url{https://doi.org/10.1073/pnas.79.8.2554},
  \DOIprefix\doi{10.1073/pnas.79.8.2554}.
%Type = Article
\bibitem[{Izhikevich(2003)}]{Izhikevich2003}
\bibinfo{author}{Izhikevich, E.}, \bibinfo{year}{2003}.
\newblock \bibinfo{title}{Simple model of spiking neurons}.
\newblock \bibinfo{journal}{{IEEE} Transactions on Neural Networks}
  \bibinfo{volume}{14}, \bibinfo{pages}{1569--1572}.
\newblock \URLprefix \url{https://doi.org/10.1109/tnn.2003.820440},
  \DOIprefix\doi{10.1109/tnn.2003.820440}.
%Type = Article
\bibitem[{Jourdain et~al.(2007)Jourdain, Bergersen, Bhaukaurally, Bezzi,
  Santello, Domercq, Matute, Tonello, Gundersen and Volterra}]{Jourdain2007}
\bibinfo{author}{Jourdain, P.}, \bibinfo{author}{Bergersen, L.H.},
  \bibinfo{author}{Bhaukaurally, K.}, \bibinfo{author}{Bezzi, P.},
  \bibinfo{author}{Santello, M.}, \bibinfo{author}{Domercq, M.},
  \bibinfo{author}{Matute, C.}, \bibinfo{author}{Tonello, F.},
  \bibinfo{author}{Gundersen, V.}, \bibinfo{author}{Volterra, A.},
  \bibinfo{year}{2007}.
\newblock \bibinfo{title}{Glutamate exocytosis from astrocytes controls
  synaptic strength}.
\newblock \bibinfo{journal}{Nature Neuroscience} \bibinfo{volume}{10},
  \bibinfo{pages}{331--339}.
\newblock \URLprefix \url{https://doi.org/10.1038/nn1849},
  \DOIprefix\doi{10.1038/nn1849}.
%Type = Article
\bibitem[{Kanakov et~al.(2019)Kanakov, Gordleeva, Ermolaeva, Jalan and
  Zaikin}]{Kanakov2019}
\bibinfo{author}{Kanakov, O.}, \bibinfo{author}{Gordleeva, S.},
  \bibinfo{author}{Ermolaeva, A.}, \bibinfo{author}{Jalan, S.},
  \bibinfo{author}{Zaikin, A.}, \bibinfo{year}{2019}.
\newblock \bibinfo{title}{Astrocyte-induced positive integrated information in
  neuron-astrocyte ensembles}.
\newblock \bibinfo{journal}{Physical Review E} \bibinfo{volume}{99}.
\newblock \URLprefix \url{https://doi.org/10.1103/physreve.99.012418},
  \DOIprefix\doi{10.1103/physreve.99.012418}.
%Type = Article
\bibitem[{Kass and Mintz(2005)}]{Kass2005}
\bibinfo{author}{Kass, J.I.}, \bibinfo{author}{Mintz, I.M.},
  \bibinfo{year}{2005}.
\newblock \bibinfo{title}{Silent plateau potentials, rhythmic bursts, and
  pacemaker firing: Three patterns of activity that coexist in quadristable
  subthalamic neurons}.
\newblock \bibinfo{journal}{Proceedings of the National Academy of Sciences}
  \bibinfo{volume}{103}, \bibinfo{pages}{183--188}.
\newblock \URLprefix \url{https://doi.org/10.1073/pnas.0506781102},
  \DOIprefix\doi{10.1073/pnas.0506781102}.
%Type = Article
\bibitem[{Kastanenka et~al.(2019)Kastanenka, Moreno-Bote, Pitt{\`{a}}, Perea,
  Eraso-Pichot, Masgrau, Poskanzer and Galea}]{Kastanenka2019}
\bibinfo{author}{Kastanenka, K.V.}, \bibinfo{author}{Moreno-Bote, R.},
  \bibinfo{author}{Pitt{\`{a}}, M.D.}, \bibinfo{author}{Perea, G.},
  \bibinfo{author}{Eraso-Pichot, A.}, \bibinfo{author}{Masgrau, R.},
  \bibinfo{author}{Poskanzer, K.E.}, \bibinfo{author}{Galea, E.},
  \bibinfo{year}{2019}.
\newblock \bibinfo{title}{A roadmap to integrate astrocytes into systems
  neuroscience}.
\newblock \bibinfo{journal}{Glia} \bibinfo{volume}{68}, \bibinfo{pages}{5--26}.
\newblock \URLprefix \url{https://doi.org/10.1002/glia.23632},
  \DOIprefix\doi{10.1002/glia.23632}.
%Type = Article
\bibitem[{Kazantsev and Asatryan(2011)}]{Kazantsev2011}
\bibinfo{author}{Kazantsev, V.B.}, \bibinfo{author}{Asatryan, S.Y.},
  \bibinfo{year}{2011}.
\newblock \bibinfo{title}{Bistability induces episodic spike communication by
  inhibitory neurons in neuronal networks}.
\newblock \bibinfo{journal}{Physical Review E} \bibinfo{volume}{84}.
\newblock \URLprefix \url{https://doi.org/10.1103/physreve.84.031913},
  \DOIprefix\doi{10.1103/physreve.84.031913}.
%Type = Article
\bibitem[{Kilpatrick et~al.(2013)Kilpatrick, Ermentrout and
  Doiron}]{Kilpatrick2013}
\bibinfo{author}{Kilpatrick, Z.P.}, \bibinfo{author}{Ermentrout, B.},
  \bibinfo{author}{Doiron, B.}, \bibinfo{year}{2013}.
\newblock \bibinfo{title}{Optimizing working memory with heterogeneity of
  recurrent cortical excitation}.
\newblock \bibinfo{journal}{Journal of Neuroscience} \bibinfo{volume}{33},
  \bibinfo{pages}{18999--19011}.
\newblock \URLprefix \url{https://doi.org/10.1523/jneurosci.1641-13.2013},
  \DOIprefix\doi{10.1523/jneurosci.1641-13.2013}.
%Type = Article
\bibitem[{Klinshov and Nekorkin(2008)}]{KLINSHOV2008}
\bibinfo{author}{Klinshov, V.V.}, \bibinfo{author}{Nekorkin, V.I.},
  \bibinfo{year}{2008}.
\newblock \bibinfo{title}{{WORKING} {MEMORY} {IN} {THE} {NETWORK} {OF}
  {NEURON}-{LIKE} {UNITS} {WITH} {NOISE}}.
\newblock \bibinfo{journal}{International Journal of Bifurcation and Chaos}
  \bibinfo{volume}{18}, \bibinfo{pages}{2743--2752}.
\newblock \URLprefix \url{https://doi.org/10.1142/s0218127408021968},
  \DOIprefix\doi{10.1142/s0218127408021968}.
%Type = Article
\bibitem[{Koulakov et~al.(2002)Koulakov, Raghavachari, Kepecs and
  Lisman}]{Koulakov2002}
\bibinfo{author}{Koulakov, A.A.}, \bibinfo{author}{Raghavachari, S.},
  \bibinfo{author}{Kepecs, A.}, \bibinfo{author}{Lisman, J.E.},
  \bibinfo{year}{2002}.
\newblock \bibinfo{title}{Model for a robust neural integrator}.
\newblock \bibinfo{journal}{Nature Neuroscience} \bibinfo{volume}{5},
  \bibinfo{pages}{775--782}.
\newblock \URLprefix \url{https://doi.org/10.1038/nn893},
  \DOIprefix\doi{10.1038/nn893}.
%Type = Article
\bibitem[{Koutsikou et~al.(2018)Koutsikou, Merrison-Hort, Buhl, Ferrario, Li,
  Borisyuk, Soffe and Roberts}]{Koutsikou2018}
\bibinfo{author}{Koutsikou, S.}, \bibinfo{author}{Merrison-Hort, R.},
  \bibinfo{author}{Buhl, E.}, \bibinfo{author}{Ferrario, A.},
  \bibinfo{author}{Li, W.C.}, \bibinfo{author}{Borisyuk, R.},
  \bibinfo{author}{Soffe, S.R.}, \bibinfo{author}{Roberts, A.},
  \bibinfo{year}{2018}.
\newblock \bibinfo{title}{A simple decision to move in response to touch
  reveals basic sensory memory and mechanisms for variable response times}.
\newblock \bibinfo{journal}{The Journal of Physiology} \bibinfo{volume}{596},
  \bibinfo{pages}{6219--6233}.
\newblock \URLprefix \url{https://doi.org/10.1113/jp276356},
  \DOIprefix\doi{10.1113/jp276356}.
%Type = Article
\bibitem[{Li and Rinzel(1994)}]{Li1994}
\bibinfo{author}{Li, Y.X.}, \bibinfo{author}{Rinzel, J.}, \bibinfo{year}{1994}.
\newblock \bibinfo{title}{Equations for {InsP}3 receptor-mediated [ca2$+$]i
  oscillations derived from a detailed kinetic model: A hodgkin-huxley like
  formalism}.
\newblock \bibinfo{journal}{Journal of Theoretical Biology}
  \bibinfo{volume}{166}, \bibinfo{pages}{461--473}.
\newblock \URLprefix \url{https://doi.org/10.1006/jtbi.1994.1041},
  \DOIprefix\doi{10.1006/jtbi.1994.1041}.
%Type = Article
\bibitem[{Lima et~al.(2014)Lima, Sardinha, Oliveira, Reis, Mota, Silva,
  Marques, Cerqueira, Pinto, Sousa and Oliveira}]{Lima2014}
\bibinfo{author}{Lima, A.}, \bibinfo{author}{Sardinha, V.M.},
  \bibinfo{author}{Oliveira, A.F.}, \bibinfo{author}{Reis, M.},
  \bibinfo{author}{Mota, C.}, \bibinfo{author}{Silva, M.A.},
  \bibinfo{author}{Marques, F.}, \bibinfo{author}{Cerqueira, J.J.},
  \bibinfo{author}{Pinto, L.}, \bibinfo{author}{Sousa, N.},
  \bibinfo{author}{Oliveira, J.F.}, \bibinfo{year}{2014}.
\newblock \bibinfo{title}{Astrocyte pathology in the prefrontal cortex impairs
  the cognitive function of rats}.
\newblock \bibinfo{journal}{Molecular Psychiatry} \bibinfo{volume}{19},
  \bibinfo{pages}{834--841}.
\newblock \URLprefix \url{https://doi.org/10.1038/mp.2013.182},
  \DOIprefix\doi{10.1038/mp.2013.182}.
%Type = Article
\bibitem[{Lisman and Idiart(1995)}]{Lisman1995}
\bibinfo{author}{Lisman, J.}, \bibinfo{author}{Idiart, M.},
  \bibinfo{year}{1995}.
\newblock \bibinfo{title}{Storage of 7 $\pm$ 2 short-term memories in
  oscillatory subcycles}.
\newblock \bibinfo{journal}{Science} \bibinfo{volume}{267},
  \bibinfo{pages}{1512--1515}.
\newblock \URLprefix \url{https://doi.org/10.1126/science.7878473},
  \DOIprefix\doi{10.1126/science.7878473}.
%Type = Article
\bibitem[{Liu et~al.(2019)Liu, Mcdaid, Harkin, Karim, Johnson, Millard, Hilder,
  Halliday, Tyrrell and Timmis}]{Liu2019}
\bibinfo{author}{Liu, J.}, \bibinfo{author}{Mcdaid, L.J.},
  \bibinfo{author}{Harkin, J.}, \bibinfo{author}{Karim, S.},
  \bibinfo{author}{Johnson, A.P.}, \bibinfo{author}{Millard, A.G.},
  \bibinfo{author}{Hilder, J.}, \bibinfo{author}{Halliday, D.M.},
  \bibinfo{author}{Tyrrell, A.M.}, \bibinfo{author}{Timmis, J.},
  \bibinfo{year}{2019}.
\newblock \bibinfo{title}{Exploring self-repair in a coupled spiking astrocyte
  neural network}.
\newblock \bibinfo{journal}{{IEEE} Transactions on Neural Networks and Learning
  Systems} \bibinfo{volume}{30}, \bibinfo{pages}{865--875}.
\newblock \URLprefix \url{https://doi.org/10.1109/tnnls.2018.2854291},
  \DOIprefix\doi{10.1109/tnnls.2018.2854291}.
%Type = Article
\bibitem[{Luca et~al.(2020)Luca, Soch, Sominsky, Nguyen, Bosakhar and
  Spencer}]{DeLuca2020}
\bibinfo{author}{Luca, S.N.D.}, \bibinfo{author}{Soch, A.},
  \bibinfo{author}{Sominsky, L.}, \bibinfo{author}{Nguyen, T.X.},
  \bibinfo{author}{Bosakhar, A.}, \bibinfo{author}{Spencer, S.J.},
  \bibinfo{year}{2020}.
\newblock \bibinfo{title}{Glial remodeling enhances short-term memory
  performance in wistar rats}.
\newblock \bibinfo{journal}{Journal of Neuroinflammation} \bibinfo{volume}{17}.
\newblock \URLprefix \url{https://doi.org/10.1186/s12974-020-1729-4},
  \DOIprefix\doi{10.1186/s12974-020-1729-4}.
%Type = Article
\bibitem[{Lundqvist et~al.(2011)Lundqvist, Herman and Lansner}]{Lundqvist2011}
\bibinfo{author}{Lundqvist, M.}, \bibinfo{author}{Herman, P.},
  \bibinfo{author}{Lansner, A.}, \bibinfo{year}{2011}.
\newblock \bibinfo{title}{Theta and gamma power increases and alpha/beta power
  decreases with memory load in an attractor network model}.
\newblock \bibinfo{journal}{Journal of Cognitive Neuroscience}
  \bibinfo{volume}{23}, \bibinfo{pages}{3008--3020}.
\newblock \DOIprefix\doi{10.1162/jocn\_a\_00029}.
%Type = Article
\bibitem[{Lundqvist et~al.(2018)Lundqvist, Herman and Miller}]{Lundqvist2018}
\bibinfo{author}{Lundqvist, M.}, \bibinfo{author}{Herman, P.},
  \bibinfo{author}{Miller, E.K.}, \bibinfo{year}{2018}.
\newblock \bibinfo{title}{Working memory: Delay activity, yes! persistent
  activity? maybe not}.
\newblock \bibinfo{journal}{The Journal of Neuroscience} \bibinfo{volume}{38},
  \bibinfo{pages}{7013--7019}.
\newblock \URLprefix \url{https://doi.org/10.1523/jneurosci.2485-17.2018},
  \DOIprefix\doi{10.1523/jneurosci.2485-17.2018}.
%Type = Article
\bibitem[{Lundqvist et~al.(2016)Lundqvist, Rose, Herman, Brincat, Buschman and
  Miller}]{Lundqvist2016}
\bibinfo{author}{Lundqvist, M.}, \bibinfo{author}{Rose, J.},
  \bibinfo{author}{Herman, P.}, \bibinfo{author}{Brincat, S.L.},
  \bibinfo{author}{Buschman, T.J.}, \bibinfo{author}{Miller, E.K.},
  \bibinfo{year}{2016}.
\newblock \bibinfo{title}{Gamma and beta bursts underlie working memory}.
\newblock \bibinfo{journal}{Neuron} \bibinfo{volume}{90},
  \bibinfo{pages}{152--164}.
\newblock \URLprefix \url{https://doi.org/10.1016/j.neuron.2016.02.028},
  \DOIprefix\doi{10.1016/j.neuron.2016.02.028}.
%Type = Article
\bibitem[{Manohar et~al.(2019)Manohar, Zokaei, Fallon, Vogels and
  Husain}]{Manohar2019}
\bibinfo{author}{Manohar, S.G.}, \bibinfo{author}{Zokaei, N.},
  \bibinfo{author}{Fallon, S.J.}, \bibinfo{author}{Vogels, T.P.},
  \bibinfo{author}{Husain, M.}, \bibinfo{year}{2019}.
\newblock \bibinfo{title}{Neural mechanisms of attending to items in working
  memory}.
\newblock \bibinfo{journal}{Neuroscience {\&} Biobehavioral Reviews}
  \bibinfo{volume}{101}, \bibinfo{pages}{1--12}.
\newblock \URLprefix \url{https://doi.org/10.1016/j.neubiorev.2019.03.017},
  \DOIprefix\doi{10.1016/j.neubiorev.2019.03.017}.
%Type = Article
\bibitem[{Mariotti et~al.(2018)Mariotti, Losi, Lia, Melone, Chiavegato,
  G{\'{o}}mez-Gonzalo, Sessolo, Bovetti, Forli, Zonta, Requie, Marcon,
  Pugliese, Viollet, Bettler, Fellin, Conti and Carmignoto}]{Mariotti2018}
\bibinfo{author}{Mariotti, L.}, \bibinfo{author}{Losi, G.},
  \bibinfo{author}{Lia, A.}, \bibinfo{author}{Melone, M.},
  \bibinfo{author}{Chiavegato, A.}, \bibinfo{author}{G{\'{o}}mez-Gonzalo, M.},
  \bibinfo{author}{Sessolo, M.}, \bibinfo{author}{Bovetti, S.},
  \bibinfo{author}{Forli, A.}, \bibinfo{author}{Zonta, M.},
  \bibinfo{author}{Requie, L.M.}, \bibinfo{author}{Marcon, I.},
  \bibinfo{author}{Pugliese, A.}, \bibinfo{author}{Viollet, C.},
  \bibinfo{author}{Bettler, B.}, \bibinfo{author}{Fellin, T.},
  \bibinfo{author}{Conti, F.}, \bibinfo{author}{Carmignoto, G.},
  \bibinfo{year}{2018}.
\newblock \bibinfo{title}{Interneuron-specific signaling evokes distinctive
  somatostatin-mediated responses in adult cortical astrocytes}.
\newblock \bibinfo{journal}{Nature Communications} \bibinfo{volume}{9}.
\newblock \URLprefix \url{https://doi.org/10.1038/s41467-017-02642-6},
  \DOIprefix\doi{10.1038/s41467-017-02642-6}.
%Type = Article
\bibitem[{Martin et~al.(2015)Martin, Bajo-Graneras, Moratalla, Perea and
  Araque}]{Martin2015}
\bibinfo{author}{Martin, R.}, \bibinfo{author}{Bajo-Graneras, R.},
  \bibinfo{author}{Moratalla, R.}, \bibinfo{author}{Perea, G.},
  \bibinfo{author}{Araque, A.}, \bibinfo{year}{2015}.
\newblock \bibinfo{title}{Circuit-specific signaling in astrocyte-neuron
  networks in basal ganglia pathways}.
\newblock \bibinfo{journal}{Science} \bibinfo{volume}{349},
  \bibinfo{pages}{730--734}.
\newblock \URLprefix \url{https://doi.org/10.1126/science.aaa7945},
  \DOIprefix\doi{10.1126/science.aaa7945}.
%Type = Article
\bibitem[{Mi et~al.(2017)Mi, Katkov and Tsodyks}]{Mi2017}
\bibinfo{author}{Mi, Y.}, \bibinfo{author}{Katkov, M.},
  \bibinfo{author}{Tsodyks, M.}, \bibinfo{year}{2017}.
\newblock \bibinfo{title}{Synaptic correlates of working memory capacity}.
\newblock \bibinfo{journal}{Neuron} \bibinfo{volume}{93},
  \bibinfo{pages}{323--330}.
\newblock \URLprefix \url{https://doi.org/10.1016/j.neuron.2016.12.004},
  \DOIprefix\doi{10.1016/j.neuron.2016.12.004}.
%Type = Article
\bibitem[{Miller et~al.(1996)Miller, Erickson and Desimone}]{Miller1996}
\bibinfo{author}{Miller, E.K.}, \bibinfo{author}{Erickson, C.A.},
  \bibinfo{author}{Desimone, R.}, \bibinfo{year}{1996}.
\newblock \bibinfo{title}{Neural mechanisms of visual working memory in
  prefrontal cortex of the macaque}.
\newblock \bibinfo{journal}{The Journal of Neuroscience} \bibinfo{volume}{16},
  \bibinfo{pages}{5154--5167}.
\newblock \URLprefix \url{https://doi.org/10.1523/jneurosci.16-16-05154.1996},
  \DOIprefix\doi{10.1523/jneurosci.16-16-05154.1996}.
%Type = Article
\bibitem[{Mongillo et~al.(2008)Mongillo, Barak and Tsodyks}]{Mongillo2008}
\bibinfo{author}{Mongillo, G.}, \bibinfo{author}{Barak, O.},
  \bibinfo{author}{Tsodyks, M.}, \bibinfo{year}{2008}.
\newblock \bibinfo{title}{Synaptic theory of working memory}.
\newblock \bibinfo{journal}{Science} \bibinfo{volume}{319},
  \bibinfo{pages}{1543--1546}.
\newblock \URLprefix \url{https://doi.org/10.1126/science.1150769},
  \DOIprefix\doi{10.1126/science.1150769}.
%Type = Article
\bibitem[{Nadkarni and Jung(2003)}]{Nadkarni2003}
\bibinfo{author}{Nadkarni, S.}, \bibinfo{author}{Jung, P.},
  \bibinfo{year}{2003}.
\newblock \bibinfo{title}{Spontaneous oscillations of dressed neurons: A new
  mechanism for epilepsy?}
\newblock \bibinfo{journal}{Physical Review Letters} \bibinfo{volume}{91}.
\newblock \URLprefix \url{https://doi.org/10.1103/physrevlett.91.268101},
  \DOIprefix\doi{10.1103/physrevlett.91.268101}.
%Type = Article
\bibitem[{Navarrete et~al.(2012)Navarrete, Perea, de~Sevilla,
  G{\'{o}}mez-Gonzalo, N{\'{u}}{\~{n}}ez, Mart{\'{\i}}n and
  Araque}]{Navarrete2012}
\bibinfo{author}{Navarrete, M.}, \bibinfo{author}{Perea, G.},
  \bibinfo{author}{de~Sevilla, D.F.}, \bibinfo{author}{G{\'{o}}mez-Gonzalo,
  M.}, \bibinfo{author}{N{\'{u}}{\~{n}}ez, A.}, \bibinfo{author}{Mart{\'{\i}}n,
  E.D.}, \bibinfo{author}{Araque, A.}, \bibinfo{year}{2012}.
\newblock \bibinfo{title}{Astrocytes mediate in vivo cholinergic-induced
  synaptic plasticity}.
\newblock \bibinfo{journal}{{PLoS} Biology} \bibinfo{volume}{10},
  \bibinfo{pages}{e1001259}.
\newblock \URLprefix \url{https://doi.org/10.1371/journal.pbio.1001259},
  \DOIprefix\doi{10.1371/journal.pbio.1001259}.
%Type = Article
\bibitem[{Nimmerjahn et~al.(2004)Nimmerjahn, Kirchhoff, Kerr and
  Helmchen}]{Nimmerjahn2004}
\bibinfo{author}{Nimmerjahn, A.}, \bibinfo{author}{Kirchhoff, F.},
  \bibinfo{author}{Kerr, J.N.D.}, \bibinfo{author}{Helmchen, F.},
  \bibinfo{year}{2004}.
\newblock \bibinfo{title}{Sulforhodamine 101 as a specific marker of astroglia
  in the neocortex in vivo}.
\newblock \bibinfo{journal}{Nature Methods} \bibinfo{volume}{1},
  \bibinfo{pages}{31--37}.
\newblock \URLprefix \url{https://doi.org/10.1038/nmeth706},
  \DOIprefix\doi{10.1038/nmeth706}.
%Type = Article
\bibitem[{Ozdemir et~al.(2020)Ozdemir, Lagler, Lagoun, Malagon-Vina,
  Laszt{\'{o}}czi and Klausberger}]{Ozdemir2020}
\bibinfo{author}{Ozdemir, A.T.}, \bibinfo{author}{Lagler, M.},
  \bibinfo{author}{Lagoun, S.}, \bibinfo{author}{Malagon-Vina, H.},
  \bibinfo{author}{Laszt{\'{o}}czi, B.}, \bibinfo{author}{Klausberger, T.},
  \bibinfo{year}{2020}.
\newblock \bibinfo{title}{Unexpected rule-changes in a working memory task
  shape the firing of histologically identified delay-tuned neurons in the
  prefrontal cortex}.
\newblock \bibinfo{journal}{Cell Reports} \bibinfo{volume}{30},
  \bibinfo{pages}{1613--1626.e4}.
\newblock \URLprefix \url{https://doi.org/10.1016/j.celrep.2019.12.102},
  \DOIprefix\doi{10.1016/j.celrep.2019.12.102}.
%Type = Article
\bibitem[{Pankratova et~al.(2019)Pankratova, Kalyakulina, Stasenko, Gordleeva,
  Lazarevich and Kazantsev}]{Pankratova2019}
\bibinfo{author}{Pankratova, E.V.}, \bibinfo{author}{Kalyakulina, A.I.},
  \bibinfo{author}{Stasenko, S.V.}, \bibinfo{author}{Gordleeva, S.Y.},
  \bibinfo{author}{Lazarevich, I.A.}, \bibinfo{author}{Kazantsev, V.B.},
  \bibinfo{year}{2019}.
\newblock \bibinfo{title}{Neuronal synchronization enhanced by
  neuron{\textendash}astrocyte interaction}.
\newblock \bibinfo{journal}{Nonlinear Dynamics} \bibinfo{volume}{97},
  \bibinfo{pages}{647--662}.
\newblock \URLprefix \url{https://doi.org/10.1007/s11071-019-05004-7},
  \DOIprefix\doi{10.1007/s11071-019-05004-7}.
%Type = Article
\bibitem[{Park et~al.(2019)Park, Bae, Kim and Jung}]{Park2019}
\bibinfo{author}{Park, J.C.}, \bibinfo{author}{Bae, J.W.},
  \bibinfo{author}{Kim, J.}, \bibinfo{author}{Jung, M.W.},
  \bibinfo{year}{2019}.
\newblock \bibinfo{title}{Dynamically changing neuronal activity supporting
  working memory for predictable and unpredictable durations}.
\newblock \bibinfo{journal}{Scientific Reports} \bibinfo{volume}{9}.
\newblock \URLprefix \url{https://doi.org/10.1038/s41598-019-52017-8},
  \DOIprefix\doi{10.1038/s41598-019-52017-8}.
%Type = Article
\bibitem[{Paukert et~al.(2014)Paukert, Agarwal, Cha, Doze, Kang and
  Bergles}]{Paukert2014}
\bibinfo{author}{Paukert, M.}, \bibinfo{author}{Agarwal, A.},
  \bibinfo{author}{Cha, J.}, \bibinfo{author}{Doze, V.A.},
  \bibinfo{author}{Kang, J.U.}, \bibinfo{author}{Bergles, D.E.},
  \bibinfo{year}{2014}.
\newblock \bibinfo{title}{Norepinephrine controls astroglial responsiveness to
  local circuit activity}.
\newblock \bibinfo{journal}{Neuron} \bibinfo{volume}{82},
  \bibinfo{pages}{1263--1270}.
\newblock \URLprefix \url{https://doi.org/10.1016/j.neuron.2014.04.038},
  \DOIprefix\doi{10.1016/j.neuron.2014.04.038}.
%Type = Article
\bibitem[{Perea and Araque(2007)}]{Perea2007}
\bibinfo{author}{Perea, G.}, \bibinfo{author}{Araque, A.},
  \bibinfo{year}{2007}.
\newblock \bibinfo{title}{Astrocytes potentiate transmitter release at single
  hippocampal synapses}.
\newblock \bibinfo{journal}{Science} \bibinfo{volume}{317},
  \bibinfo{pages}{1083--1086}.
\newblock \URLprefix \url{https://doi.org/10.1126/science.1144640},
  \DOIprefix\doi{10.1126/science.1144640}.
%Type = Article
\bibitem[{Perea et~al.(2014)Perea, Yang, Boyden and Sur}]{Perea2014}
\bibinfo{author}{Perea, G.}, \bibinfo{author}{Yang, A.},
  \bibinfo{author}{Boyden, E.S.}, \bibinfo{author}{Sur, M.},
  \bibinfo{year}{2014}.
\newblock \bibinfo{title}{Optogenetic astrocyte activation modulates response
  selectivity of visual cortex neurons in vivo}.
\newblock \bibinfo{journal}{Nature Communications} \bibinfo{volume}{5}.
\newblock \URLprefix \url{https://doi.org/10.1038/ncomms4262},
  \DOIprefix\doi{10.1038/ncomms4262}.
%Type = Article
\bibitem[{Pitt{\`{a}} et~al.(2016)Pitt{\`{a}}, Brunel and
  Volterra}]{DePitt2016}
\bibinfo{author}{Pitt{\`{a}}, M.D.}, \bibinfo{author}{Brunel, N.},
  \bibinfo{author}{Volterra, A.}, \bibinfo{year}{2016}.
\newblock \bibinfo{title}{Astrocytes: Orchestrating synaptic plasticity?}
\newblock \bibinfo{journal}{Neuroscience} \bibinfo{volume}{323},
  \bibinfo{pages}{43--61}.
\newblock \URLprefix \url{https://doi.org/10.1016/j.neuroscience.2015.04.001},
  \DOIprefix\doi{10.1016/j.neuroscience.2015.04.001}.
%Type = Article
\bibitem[{Poskanzer and Yuste(2016)}]{Poskanzer2016}
\bibinfo{author}{Poskanzer, K.E.}, \bibinfo{author}{Yuste, R.},
  \bibinfo{year}{2016}.
\newblock \bibinfo{title}{Astrocytes regulate cortical state switching in
  vivo}.
\newblock \bibinfo{journal}{Proceedings of the National Academy of Sciences}
  \bibinfo{volume}{113}, \bibinfo{pages}{E2675--E2684}.
\newblock \URLprefix \url{https://doi.org/10.1073/pnas.1520759113},
  \DOIprefix\doi{10.1073/pnas.1520759113}.
%Type = Article
\bibitem[{Robin et~al.(2018)Robin, da~Cruz, Langlais, Martin-Fernandez,
  Metna-Laurent, Busquets-Garcia, Bellocchio, Soria-Gomez, Papouin, Varilh,
  Sherwood, Belluomo, Balcells, Matias, Bosier, Drago, Eeckhaut, Smolders,
  Georges, Araque, Panatier, Oliet and Marsicano}]{Robin2018}
\bibinfo{author}{Robin, L.M.}, \bibinfo{author}{da~Cruz, J.F.O.},
  \bibinfo{author}{Langlais, V.C.}, \bibinfo{author}{Martin-Fernandez, M.},
  \bibinfo{author}{Metna-Laurent, M.}, \bibinfo{author}{Busquets-Garcia, A.},
  \bibinfo{author}{Bellocchio, L.}, \bibinfo{author}{Soria-Gomez, E.},
  \bibinfo{author}{Papouin, T.}, \bibinfo{author}{Varilh, M.},
  \bibinfo{author}{Sherwood, M.W.}, \bibinfo{author}{Belluomo, I.},
  \bibinfo{author}{Balcells, G.}, \bibinfo{author}{Matias, I.},
  \bibinfo{author}{Bosier, B.}, \bibinfo{author}{Drago, F.},
  \bibinfo{author}{Eeckhaut, A.V.}, \bibinfo{author}{Smolders, I.},
  \bibinfo{author}{Georges, F.}, \bibinfo{author}{Araque, A.},
  \bibinfo{author}{Panatier, A.}, \bibinfo{author}{Oliet, S.H.},
  \bibinfo{author}{Marsicano, G.}, \bibinfo{year}{2018}.
\newblock \bibinfo{title}{Astroglial {CB}1 receptors determine synaptic
  d-serine availability to enable recognition memory}.
\newblock \bibinfo{journal}{Neuron} \bibinfo{volume}{98},
  \bibinfo{pages}{935--944.e5}.
\newblock \URLprefix \url{https://doi.org/10.1016/j.neuron.2018.04.034},
  \DOIprefix\doi{10.1016/j.neuron.2018.04.034}.
%Type = Article
\bibitem[{Runyan et~al.(2017)Runyan, Piasini, Panzeri and Harvey}]{Runyan2017}
\bibinfo{author}{Runyan, C.A.}, \bibinfo{author}{Piasini, E.},
  \bibinfo{author}{Panzeri, S.}, \bibinfo{author}{Harvey, C.D.},
  \bibinfo{year}{2017}.
\newblock \bibinfo{title}{Distinct timescales of population coding across
  cortex}.
\newblock \bibinfo{journal}{Nature} \bibinfo{volume}{548},
  \bibinfo{pages}{92--96}.
\newblock \URLprefix \url{https://doi.org/10.1038/nature23020},
  \DOIprefix\doi{10.1038/nature23020}.
%Type = Article
\bibitem[{Sandberg et~al.(2003)Sandberg, Tegn{\'{e}}r and
  Lansner}]{Sandberg2003}
\bibinfo{author}{Sandberg, A.}, \bibinfo{author}{Tegn{\'{e}}r, J.},
  \bibinfo{author}{Lansner, A.}, \bibinfo{year}{2003}.
\newblock \bibinfo{title}{A working memory model based on fast hebbian
  learning}.
\newblock \bibinfo{journal}{Network: Computation in Neural Systems}
  \bibinfo{volume}{14}, \bibinfo{pages}{789--802}.
\newblock \URLprefix \url{https://doi.org/10.1088/0954-898x\_14\_4\_309},
  \DOIprefix\doi{10.1088/0954-898x\_14\_4\_309}.
%Type = Article
\bibitem[{Santello et~al.(2019)Santello, Toni and Volterra}]{Santello2019}
\bibinfo{author}{Santello, M.}, \bibinfo{author}{Toni, N.},
  \bibinfo{author}{Volterra, A.}, \bibinfo{year}{2019}.
\newblock \bibinfo{title}{Astrocyte function from information processing to
  cognition and cognitive impairment}.
\newblock \bibinfo{journal}{Nature Neuroscience} \bibinfo{volume}{22},
  \bibinfo{pages}{154--166}.
\newblock \URLprefix \url{https://doi.org/10.1038/s41593-018-0325-8},
  \DOIprefix\doi{10.1038/s41593-018-0325-8}.
%Type = Article
\bibitem[{Sardinha et~al.(2017)Sardinha, Guerra-Gomes, Caetano, Tavares,
  Martins, Reis, Correia, Teixeira-Castro, Pinto, Sousa and
  Oliveira}]{Sardinha2017}
\bibinfo{author}{Sardinha, V.M.}, \bibinfo{author}{Guerra-Gomes, S.},
  \bibinfo{author}{Caetano, I.}, \bibinfo{author}{Tavares, G.},
  \bibinfo{author}{Martins, M.}, \bibinfo{author}{Reis, J.S.},
  \bibinfo{author}{Correia, J.S.}, \bibinfo{author}{Teixeira-Castro, A.},
  \bibinfo{author}{Pinto, L.}, \bibinfo{author}{Sousa, N.},
  \bibinfo{author}{Oliveira, J.F.}, \bibinfo{year}{2017}.
\newblock \bibinfo{title}{Astrocytic signaling supports hippocampal-prefrontal
  theta synchronization and cognitive function}.
\newblock \bibinfo{journal}{Glia} \bibinfo{volume}{65},
  \bibinfo{pages}{1944--1960}.
\newblock \URLprefix \url{https://doi.org/10.1002/glia.23205},
  \DOIprefix\doi{10.1002/glia.23205}.
%Type = Article
\bibitem[{Sasaki et~al.(2011)Sasaki, Kuga, Namiki, Matsuki and
  Ikegaya}]{Sasaki2011}
\bibinfo{author}{Sasaki, T.}, \bibinfo{author}{Kuga, N.},
  \bibinfo{author}{Namiki, S.}, \bibinfo{author}{Matsuki, N.},
  \bibinfo{author}{Ikegaya, Y.}, \bibinfo{year}{2011}.
\newblock \bibinfo{title}{Locally synchronized astrocytes}.
\newblock \bibinfo{journal}{Cerebral Cortex} \bibinfo{volume}{21},
  \bibinfo{pages}{1889--1900}.
\newblock \URLprefix \url{https://doi.org/10.1093/cercor/bhq256},
  \DOIprefix\doi{10.1093/cercor/bhq256}.
%Type = Article
\bibitem[{Savtchouk and Volterra(2018)}]{Savtchouk2018}
\bibinfo{author}{Savtchouk, I.}, \bibinfo{author}{Volterra, A.},
  \bibinfo{year}{2018}.
\newblock \bibinfo{title}{Gliotransmission: Beyond black-and-white}.
\newblock \bibinfo{journal}{The Journal of Neuroscience} \bibinfo{volume}{38},
  \bibinfo{pages}{14--25}.
\newblock \URLprefix \url{https://doi.org/10.1523/jneurosci.0017-17.2017},
  \DOIprefix\doi{10.1523/jneurosci.0017-17.2017}.
%Type = Article
\bibitem[{Schummers et~al.(2008)Schummers, Yu and Sur}]{Schummers2008}
\bibinfo{author}{Schummers, J.}, \bibinfo{author}{Yu, H.},
  \bibinfo{author}{Sur, M.}, \bibinfo{year}{2008}.
\newblock \bibinfo{title}{Tuned responses of astrocytes and their influence on
  hemodynamic signals in the visual cortex}.
\newblock \bibinfo{journal}{Science} \bibinfo{volume}{320},
  \bibinfo{pages}{1638--1643}.
\newblock \URLprefix \url{https://doi.org/10.1126/science.1156120},
  \DOIprefix\doi{10.1126/science.1156120}.
%Type = Article
\bibitem[{Shafi et~al.(2007)Shafi, Zhou, Quintana, Chow, Fuster and
  Bodner}]{Shafi2007}
\bibinfo{author}{Shafi, M.}, \bibinfo{author}{Zhou, Y.},
  \bibinfo{author}{Quintana, J.}, \bibinfo{author}{Chow, C.},
  \bibinfo{author}{Fuster, J.}, \bibinfo{author}{Bodner, M.},
  \bibinfo{year}{2007}.
\newblock \bibinfo{title}{Variability in neuronal activity in primate cortex
  during working memory tasks}.
\newblock \bibinfo{journal}{Neuroscience} \bibinfo{volume}{146},
  \bibinfo{pages}{1082--1108}.
\newblock \URLprefix \url{https://doi.org/10.1016/j.neuroscience.2006.12.072},
  \DOIprefix\doi{10.1016/j.neuroscience.2006.12.072}.
%Type = Article
\bibitem[{Shen and Wilde(2007)}]{Shen2007}
\bibinfo{author}{Shen, X.}, \bibinfo{author}{Wilde, P.D.},
  \bibinfo{year}{2007}.
\newblock \bibinfo{title}{Long-term neuronal behavior caused by two synaptic
  modification mechanisms}.
\newblock \bibinfo{journal}{Neurocomputing} \bibinfo{volume}{70},
  \bibinfo{pages}{1482--1488}.
\newblock \URLprefix \url{https://doi.org/10.1016/j.neucom.2006.05.011},
  \DOIprefix\doi{10.1016/j.neucom.2006.05.011}.
%Type = Article
\bibitem[{Sreenivasan and D'Esposito(2019)}]{Sreenivasan2019}
\bibinfo{author}{Sreenivasan, K.K.}, \bibinfo{author}{D'Esposito, M.},
  \bibinfo{year}{2019}.
\newblock \bibinfo{title}{The what, where and how of delay activity}.
\newblock \bibinfo{journal}{Nature Reviews Neuroscience} \bibinfo{volume}{20},
  \bibinfo{pages}{466--481}.
\newblock \URLprefix \url{https://doi.org/10.1038/s41583-019-0176-7},
  \DOIprefix\doi{10.1038/s41583-019-0176-7}.
%Type = Article
\bibitem[{Stellwagen and Malenka(2006)}]{Stellwagen2006}
\bibinfo{author}{Stellwagen, D.}, \bibinfo{author}{Malenka, R.C.},
  \bibinfo{year}{2006}.
\newblock \bibinfo{title}{Synaptic scaling mediated by glial {TNF}-$\alpha$}.
\newblock \bibinfo{journal}{Nature} \bibinfo{volume}{440},
  \bibinfo{pages}{1054--1059}.
\newblock \URLprefix \url{https://doi.org/10.1038/nature04671},
  \DOIprefix\doi{10.1038/nature04671}.
%Type = Article
\bibitem[{Stobart et~al.(2018)Stobart, Ferrari, Barrett, Gl\"{u}ck, Stobart,
  Zuend and Weber}]{Stobart2018}
\bibinfo{author}{Stobart, J.L.}, \bibinfo{author}{Ferrari, K.D.},
  \bibinfo{author}{Barrett, M.J.}, \bibinfo{author}{Gl\"{u}ck, C.},
  \bibinfo{author}{Stobart, M.J.}, \bibinfo{author}{Zuend, M.},
  \bibinfo{author}{Weber, B.}, \bibinfo{year}{2018}.
\newblock \bibinfo{title}{Cortical circuit activity evokes rapid astrocyte
  calcium signals on a similar timescale to neurons}.
\newblock \bibinfo{journal}{Neuron} \bibinfo{volume}{98},
  \bibinfo{pages}{726--735.e4}.
\newblock \URLprefix \url{https://doi.org/10.1016/j.neuron.2018.03.050},
  \DOIprefix\doi{10.1016/j.neuron.2018.03.050}.
%Type = Article
\bibitem[{Stokes et~al.(2013)Stokes, Kusunoki, Sigala, Nili, Gaffan and
  Duncan}]{Stokes2013}
\bibinfo{author}{Stokes, M.G.}, \bibinfo{author}{Kusunoki, M.},
  \bibinfo{author}{Sigala, N.}, \bibinfo{author}{Nili, H.},
  \bibinfo{author}{Gaffan, D.}, \bibinfo{author}{Duncan, J.},
  \bibinfo{year}{2013}.
\newblock \bibinfo{title}{Dynamic coding for cognitive control in prefrontal
  cortex}.
\newblock \bibinfo{journal}{Neuron} \bibinfo{volume}{78},
  \bibinfo{pages}{364--375}.
\newblock \URLprefix \url{https://doi.org/10.1016/j.neuron.2013.01.039},
  \DOIprefix\doi{10.1016/j.neuron.2013.01.039}.
%Type = Article
\bibitem[{Takata and Hirase(2008)}]{Takata2008}
\bibinfo{author}{Takata, N.}, \bibinfo{author}{Hirase, H.},
  \bibinfo{year}{2008}.
\newblock \bibinfo{title}{Cortical layer 1 and layer 2/3 astrocytes exhibit
  distinct calcium dynamics in vivo}.
\newblock \bibinfo{journal}{{PLoS} {ONE}} \bibinfo{volume}{3},
  \bibinfo{pages}{e2525}.
\newblock \URLprefix \url{https://doi.org/10.1371/journal.pone.0002525},
  \DOIprefix\doi{10.1371/journal.pone.0002525}.
%Type = Article
\bibitem[{Takata et~al.(2011)Takata, Mishima, Hisatsune, Nagai, Ebisui,
  Mikoshiba and Hirase}]{Takata2011}
\bibinfo{author}{Takata, N.}, \bibinfo{author}{Mishima, T.},
  \bibinfo{author}{Hisatsune, C.}, \bibinfo{author}{Nagai, T.},
  \bibinfo{author}{Ebisui, E.}, \bibinfo{author}{Mikoshiba, K.},
  \bibinfo{author}{Hirase, H.}, \bibinfo{year}{2011}.
\newblock \bibinfo{title}{Astrocyte calcium signaling transforms cholinergic
  modulation to cortical plasticity in vivo}.
\newblock \bibinfo{journal}{Journal of Neuroscience} \bibinfo{volume}{31},
  \bibinfo{pages}{18155--18165}.
\newblock \URLprefix \url{https://doi.org/10.1523/jneurosci.5289-11.2011},
  \DOIprefix\doi{10.1523/jneurosci.5289-11.2011}.
%Type = Article
\bibitem[{Tewari and Parpura(2013)}]{Tewari2013}
\bibinfo{author}{Tewari, S.}, \bibinfo{author}{Parpura, V.},
  \bibinfo{year}{2013}.
\newblock \bibinfo{title}{A possible role of astrocytes in contextual memory
  retrieval: An analysis obtained using a quantitative framework}.
\newblock \bibinfo{journal}{Frontiers in Computational Neuroscience}
  \bibinfo{volume}{7}.
\newblock \URLprefix \url{https://doi.org/10.3389/fncom.2013.00145},
  \DOIprefix\doi{10.3389/fncom.2013.00145}.
%Type = Article
\bibitem[{Tsodyks and Markram(1997)}]{Tsodyks1997}
\bibinfo{author}{Tsodyks, M.V.}, \bibinfo{author}{Markram, H.},
  \bibinfo{year}{1997}.
\newblock \bibinfo{title}{The neural code between neocortical pyramidal neurons
  depends on neurotransmitter release probability}.
\newblock \bibinfo{journal}{Proceedings of the National Academy of Sciences}
  \bibinfo{volume}{94}, \bibinfo{pages}{719--723}.
\newblock \URLprefix \url{https://doi.org/10.1073/pnas.94.2.719},
  \DOIprefix\doi{10.1073/pnas.94.2.719}.
%Type = Article
\bibitem[{Ullah et~al.(2006)Ullah, Jung and Cornell-Bell}]{ULLAH2006}
\bibinfo{author}{Ullah, G.}, \bibinfo{author}{Jung, P.},
  \bibinfo{author}{Cornell-Bell, A.}, \bibinfo{year}{2006}.
\newblock \bibinfo{title}{Anti-phase calcium oscillations in astrocytes via
  inositol (1, 4, 5)-trisphosphate regeneration}.
\newblock \bibinfo{journal}{Cell Calcium} \bibinfo{volume}{39},
  \bibinfo{pages}{197--208}.
\newblock \URLprefix \url{https://doi.org/10.1016/j.ceca.2005.10.009},
  \DOIprefix\doi{10.1016/j.ceca.2005.10.009}.
%Type = Article
\bibitem[{Wade et~al.(2011)Wade, McDaid, Harkin, Crunelli and Kelso}]{Wade2011}
\bibinfo{author}{Wade, J.J.}, \bibinfo{author}{McDaid, L.J.},
  \bibinfo{author}{Harkin, J.}, \bibinfo{author}{Crunelli, V.},
  \bibinfo{author}{Kelso, J.A.S.}, \bibinfo{year}{2011}.
\newblock \bibinfo{title}{Bidirectional coupling between astrocytes and neurons
  mediates learning and dynamic coordination in the brain: A multiple modeling
  approach}.
\newblock \bibinfo{journal}{{PLoS} {ONE}} \bibinfo{volume}{6},
  \bibinfo{pages}{e29445}.
\newblock \URLprefix \url{https://doi.org/10.1371/journal.pone.0029445},
  \DOIprefix\doi{10.1371/journal.pone.0029445}.
%Type = Article
\bibitem[{Wang et~al.(2006a)Wang, Lou, Xu, Tian, Peng, Han, Kang, Takano and
  Nedergaard}]{Wang2006_2}
\bibinfo{author}{Wang, X.}, \bibinfo{author}{Lou, N.}, \bibinfo{author}{Xu,
  Q.}, \bibinfo{author}{Tian, G.F.}, \bibinfo{author}{Peng, W.G.},
  \bibinfo{author}{Han, X.}, \bibinfo{author}{Kang, J.},
  \bibinfo{author}{Takano, T.}, \bibinfo{author}{Nedergaard, M.},
  \bibinfo{year}{2006}a.
\newblock \bibinfo{title}{Astrocytic ca2$+$ signaling evoked by sensory
  stimulation in vivo}.
\newblock \bibinfo{journal}{Nature Neuroscience} \bibinfo{volume}{9},
  \bibinfo{pages}{816--823}.
\newblock \URLprefix \url{https://doi.org/10.1038/nn1703},
  \DOIprefix\doi{10.1038/nn1703}.
%Type = Article
\bibitem[{Wang(2001)}]{Wang2001}
\bibinfo{author}{Wang, X.J.}, \bibinfo{year}{2001}.
\newblock \bibinfo{title}{Synaptic reverberation underlying mnemonic persistent
  activity}.
\newblock \bibinfo{journal}{Trends in Neurosciences} \bibinfo{volume}{24},
  \bibinfo{pages}{455--463}.
\newblock \URLprefix \url{https://doi.org/10.1016/s0166-2236(00)01868-3},
  \DOIprefix\doi{10.1016/s0166-2236(00)01868-3}.
%Type = Article
\bibitem[{Wang et~al.(2006b)Wang, Markram, Goodman, Berger, Ma and
  Goldman-Rakic}]{Wang2006}
\bibinfo{author}{Wang, Y.}, \bibinfo{author}{Markram, H.},
  \bibinfo{author}{Goodman, P.H.}, \bibinfo{author}{Berger, T.K.},
  \bibinfo{author}{Ma, J.}, \bibinfo{author}{Goldman-Rakic, P.S.},
  \bibinfo{year}{2006}b.
\newblock \bibinfo{title}{Heterogeneity in the pyramidal network of the medial
  prefrontal cortex}.
\newblock \bibinfo{journal}{Nature Neuroscience} \bibinfo{volume}{9},
  \bibinfo{pages}{534--542}.
\newblock \URLprefix \url{https://doi.org/10.1038/nn1670},
  \DOIprefix\doi{10.1038/nn1670}.
%Type = Article
\bibitem[{Wimmer et~al.(2014)Wimmer, Nykamp, Constantinidis and
  Compte}]{Wimmer2014}
\bibinfo{author}{Wimmer, K.}, \bibinfo{author}{Nykamp, D.Q.},
  \bibinfo{author}{Constantinidis, C.}, \bibinfo{author}{Compte, A.},
  \bibinfo{year}{2014}.
\newblock \bibinfo{title}{Bump attractor dynamics in prefrontal cortex explains
  behavioral precision in spatial working memory}.
\newblock \bibinfo{journal}{Nature Neuroscience} \bibinfo{volume}{17},
  \bibinfo{pages}{431--439}.
\newblock \URLprefix \url{https://doi.org/10.1038/nn.3645},
  \DOIprefix\doi{10.1038/nn.3645}.
%Type = Article
\bibitem[{Wu et~al.(2018)Wu, Gordleeva, Tang, Shih, Dembitskaya and
  Semyanov}]{Wu2018}
\bibinfo{author}{Wu, Y.W.}, \bibinfo{author}{Gordleeva, S.},
  \bibinfo{author}{Tang, X.}, \bibinfo{author}{Shih, P.Y.},
  \bibinfo{author}{Dembitskaya, Y.}, \bibinfo{author}{Semyanov, A.},
  \bibinfo{year}{2018}.
\newblock \bibinfo{title}{Morphological profile determines the frequency of
  spontaneous calcium events in astrocytic processes}.
\newblock \bibinfo{journal}{Glia} \bibinfo{volume}{67},
  \bibinfo{pages}{246--262}.
\newblock \URLprefix \url{https://doi.org/10.1002/glia.23537},
  \DOIprefix\doi{10.1002/glia.23537}.
%Type = Article
\bibitem[{Zylberberg and Strowbridge(2017)}]{Zylberberg2017}
\bibinfo{author}{Zylberberg, J.}, \bibinfo{author}{Strowbridge, B.W.},
  \bibinfo{year}{2017}.
\newblock \bibinfo{title}{Mechanisms of persistent activity in cortical
  circuits: Possible neural substrates for working memory}.
\newblock \bibinfo{journal}{Annual Review of Neuroscience}
  \bibinfo{volume}{40}, \bibinfo{pages}{603--627}.
\newblock \URLprefix \url{https://doi.org/10.1146/annurev-neuro-070815-014006},
  \DOIprefix\doi{10.1146/annurev-neuro-070815-014006}.

\end{thebibliography}





%%\bibitem[ ()]{}

%%\end{thebibliography}
\end{document}